\documentclass[conference]{IEEEtran}

\usepackage{graphicx}
	\setkeys{Gin}{width=\textwidth, height=\textheight, keepaspectratio}
\graphicspath{{images/}}

\usepackage{algorithm}
\usepackage[hyphens]{url}
\usepackage{algpseudocode}
\usepackage{amsthm}
\usepackage{amsmath}
\usepackage{xspace}
\usepackage{array}

\usepackage{bbold}
\usepackage{paralist}
\usepackage{xcolor}
\usepackage[small,bf]{caption}
\usepackage[skip=0pt]{subcaption}
\usepackage{booktabs}
\usepackage[square,sort,comma,numbers,compress]{natbib}
\usepackage{amssymb}
\usepackage{pifont}
\usepackage{multirow}
\usepackage{listings}

\definecolor{editorGray}{rgb}{0.95, 0.95, 0.95}
\definecolor{editorGreen}{rgb}{0,0.6,0}
\definecolor{brown}{rgb}{0.69,0.31,0.31}
\definecolor{gray}{rgb}{0.5,0.5,0.5}

\lstdefinelanguage{JavaScript}{
  morekeywords={typeof, window, new, true, false, catch, function, return, null, catch, switch, var, if, in, else, case, break, document, write, createElement, width, height, display, visibility, border, document.write},
  morecomment=[s]{/*}{*/},
  morecomment=[l]//,
  morestring=[b]",
  morestring=[b]'
}

\lstdefinestyle{htmlcssjs} {%
  backgroundcolor=\color{editorGray},
  basicstyle=\fontsize{8}{8}\fontencoding{T1}\ttfamily,
  frame=tb,
  captionpos=b,
  belowcaptionskip=\medskipamount,
  xleftmargin={0.5cm},
  numbers=left,
  stepnumber=1,
  firstnumber=1,
  numberfirstline=true,	
  identifierstyle=\color{black},
  keywordstyle=\color{blue}\ttfamily,%\bfseries,
  ndkeywordstyle=\color{editorGreen}\ttfamily,
  stringstyle=\color{black}\ttfamily,
  commentstyle=\color{brown}\ttfamily,
  language=JavaScript,
  alsodigit={.:;},	
  tabsize=2,
  showtabs=false,
  showspaces=false,
  showstringspaces=false,
  extendedchars=true,
  breaklines=true,
  numberstyle=\tiny\color{gray},
  literate=%
  {Ö}{{\"O}}1
  {Ä}{{\"A}}1
  {Ü}{{\"U}}1
  {ß}{{\ss}}1
  {ü}{{\"u}}1
  {ä}{{\"a}}1
  {ö}{{\"o}}1
}

\usepackage[marginal,hang,stable]{footmisc}
\renewcommand{\footnoterule}{
	\kern -3pt
	\hrule width 1in
	\kern 2pt
}

\usepackage[bookmarks=true, bookmarksnumbered=true, colorlinks=true, linkcolor=linkcol, citecolor=citecol, urlcolor=urlcol, hypertexnames=true]{hyperref}
\definecolor{darkgreen}{RGB}{0, 100, 0}
\definecolor{linkcol}{rgb}{0.3,0,0}
\definecolor{citecol}{rgb}{0.3,0,0}
\definecolor{urlcol}{rgb}{0.3,0,0}
\definecolor{vlightgray}{gray}{0.925}

\usepackage[scaled=0.8]{FiraMono}

\theoremstyle{definition}
\newtheorem{definition}{Definition}

\newcommand{\SYSNAME}{FP-Fed\xspace}
\newcommand{\SYSFULL}{Browser \underline{F}inger\underline{p}rint Detection via \underline{Fed}erated Learning\xspace}
\newcommand{\HEP}{\textit{Ext High Entropy}\xspace}
\newcommand{\HEPFull}{Extended High Entropy\xspace}

\newcommand{\descrit}[1]{\vspace{0.02cm}\noindent\textbf{\em #1}}

\newcommand{\descr}[1]{\smallskip\noindent\textbf{#1}}
\newcommand{\mywidth}{0.80}
\newcommand{\reduce}{\vspace{0.0cm}}
\begin{document}

	\title{\SYSNAME: Privacy-Preserving Federated Detection of Browser Fingerprinting$^*$\vspace*{-0.3cm}}
  \author{\IEEEauthorblockN{Meenatchi Sundaram Muthu Selva Annamalai}
  \IEEEauthorblockA{University College London\\
  meenatchi.annamalai.22@ucl.ac.uk}
  \and
  \IEEEauthorblockN{Igor Bilogrevic}
  \IEEEauthorblockA{Google\\
  ibilogrevic@google.com}
  \and
  \IEEEauthorblockN{Emiliano De Cristofaro}
  \IEEEauthorblockA{University of California, Riverside\\
  emilianodc@cs.ucr.edu}}

\maketitle

\renewcommand*{\thefootnote}{\fnsymbol{footnote}}
  
\footnotetext{$^*$Published in the Proceedings of the 31st Network and Distributed System Security Symposium (NDSS 2024), please cite accordingly.}

\begin{abstract}
Browser fingerprinting often provides an attractive alternative to third-party cookies for tracking users across the web. 
In fact, the increasing restrictions on third-party cookies placed by common web browsers and recent regulations like the GDPR may accelerate the transition.
To counter browser fingerprinting, previous work proposed several techniques to detect its prevalence and severity. 
However, these rely on 1) centralized web crawls and/or 2) computationally intensive operations to extract and process signals (e.g., information-flow and static analysis). 

To address these limitations, we present \SYSNAME, the first distributed system for browser fingerprinting detection.
Using \SYSNAME, users can collaboratively train on-device models based on their real browsing patterns, without sharing their training data with a central entity, by relying on Differentially Private Federated Learning (DP-FL).
To demonstrate its feasibility and effectiveness, we evaluate \SYSNAME's performance on a set of 18.3k popular websites with different privacy levels, numbers of participants, and features extracted from the scripts. 
Our experiments show that \SYSNAME achieves reasonably high detection performance and can perform both training and inference efficiently, on-device, by only relying on runtime signals extracted from the execution trace, without requiring any resource-intensive operation.
\end{abstract}

\renewcommand{\thefootnote}{\arabic{footnote}} 

\section{Introduction}
\label{sec:intro}
As users browse the web, they are often tracked across multiple unrelated websites by means of third-party cookies.
Although this type of tracking can have valid use cases (e.g., ad conversion~\cite{krishnamurthy2009privacy}), it is generally considered a privacy violation~\cite{thirdpartycookies}.
To protect users, browsers like Safari~\cite{safari2020}, Firefox~\cite{firefox2019}, and Brave~\cite{brave2022} have blocked or restricted third-party cookies by default. 
Chrome, the most popular web browser, will also start deprecating them in 2024~\cite{chrome2023}.

To circumvent these restrictions, trackers are turning to alternative methods like bounce tracking~\cite{bounce2020} and browser fingerprinting~\cite{fingerprinting2019}.
The latter %
uses client-side information to build unique user identifiers, %
typically via Javascript programs gathering device information, e.g., screen resolution, installed fonts, etc.~\cite{englehardt2016online}.
This is then combined and hashed to generate a unique identifier for the user's browser, which remains stable over time regardless of the websites visited~\cite{pugliese2020long}. 

While browser fingerprinting can be used for acceptable purposes, e.g., web authentication~\cite{alaca2016device,laperdrix2019morellian}, bot~\cite{bursztein2016picasso,fingerprintpro,cloudflarebot} or fraud detection~\cite{iovationfraud2019,relix2018}, it largely represents a threat to user privacy~\cite{safaribrowserfp2018,firefoxbrowserfp2020,w3cbrowserfp,privacyBudget2022}.
In fact, it can be even more intrusive than third-party cookies: the latter are easily detectable and can be cleared at any time, 
whereas browser fingerprinting is less transparent, and countermeasures often result in significant website breakage~\cite{iqbal2021fingerprinting,amjad2023blocking}.
Moreover, it can be effective even in incognito mode~\cite{akhavani2021browserprint} and potentially track users for months~\cite{pugliese2020long}.
Domains using browser fingerprinting have increased in recent years -- from 519 in the top 1M websites in 2016 to 2,349 in the top 100k websites in 2019~\cite{iqbal2021fingerprinting}. %
Also considering that large-scale cross-site tracking will likely move away from cookie-based tracking, it is fair to assume fingerprinting will likely continue to grow in prevalence and severity. 

Early research on browser fingerprinting facilitated the creation and manual curation of blocklists~\cite{disconnect2018,easylist,easyprivacy,privacybadger} and basic heuristics~\cite{acar2014web,englehardt2016online}.
More recently, machine learning has been used to build fingerprinting detectors with high precision and recall~\cite{ikram2017towards,das2018web,iqbal2021fingerprinting}.
These methods rely on one entity performing a large-scale crawl of the web (typically, top-ranked websites) to collect scripts, which are then labeled and used to train detection models.
These techniques showcase the feasibility of detecting browser fingerprinting with machine learning-based approaches.

\descr{Limitations of Centralized Approaches.} 
Alas, %
crawlers can rarely replicate human-like browsing behavior and interactions with a site, e.g., they are often identified by bot detectors, cannot operate beyond login and paywalls, or solve CAPTCHAs~\cite{acar2013fpdetective}.
Moreover, the scripts' behavior may differ compared to real-world interactions, depending on the type of device, OS, etc.
While it is possible in principle for a crawler to simulate these attributes (e.g., using different user-agent headers), in practice, it might be hard to do it correctly and, perhaps more importantly, to extensively cover the range of different devices, OS, etc.
To this end, we also conduct a small-scale study on the top 300 domains from the Tranco ranking list and find that crawls involving real users (logging in, solving CAPTCHAs, etc.) can capture 3 times more fingerprinting scripts than automated/centralized ones proposed in previous work~\cite{englehardt2016online,iqbal2021fingerprinting}.
In Appendix~\ref{app:missed_scripts}, we provide an example of a script missed by an automated crawler but captured by user interaction.

Overall, training data built from centralized crawlers may fail to visit a non-negligible number of (potentially fingerprinting) websites, including top-ranked ones~\cite{iqbal2021fingerprinting}.
One other option could be gathering real-world observations from different users as they browse various websites; however, data collected from websites might reveal sensitive information such as medical conditions~\cite{nhsfb}, which could affect users' privacy.
Alternatively, each user could collect their own data and train a detection model locally; while providing optimal privacy, this approach is unlikely to provide meaningful accuracy.

\descr{Federated Learning and its Challenges.} As a result, we opt to build on Federated Learning (FL)~\cite{mcmahan2016federated}, a collaborative learning approach seeking a reasonable privacy-utility compromise between local-only and fully centralized training.
With FL, users train models locally but collaborate to build a global model, which lets them acquire knowledge from other users' data by only sharing (less sensitive) model updates.

However, it is not trivial to federate high-precision, high-recall classification algorithms, and in particular, to do so in an efficient, scalable manner, e.g., adapting to the federated setting existing classifiers~\cite{iqbal2021fingerprinting} that rely on thousands of complex, hand-crafted, features extracted from each script.
Also, executing complex algorithms may severely impact the browser's performance, especially when deploying to clients with a wide range of computational power and resource constraints.

Moreover, although sharing model updates rather than raw data is inherently less privacy-invasive, inference attacks are still possible that allow adversaries to learn sensitive information about users' training data~\cite{melis2019exploiting,nasr2019comprehensive}.
Therefore, it is imperative to provide rigorous privacy guarantees through the Differential Privacy framework~\cite{dwork2014algorithmic}.
Alas, this is likely to introduce a loss in model performance and thus has to be done in a way that preserves a reasonable privacy-utility tradeoff. 

\descr{Contributions.} 
This paper introduces \SYSNAME (\SYSFULL) -- to the best of our knowledge, the first distributed system for detecting fingerprinting in the wild. 
\SYSNAME relies on Differentially Private Federated Learning (DP-FL) and achieves reasonably high accuracy, with minimal false positives, while providing formal privacy guarantees. 
To assess its feasibility and explore its deployment challenges, we analyze the performance of \SYSNAME along several axes, including different levels of privacy, number of participants, and feature sets. %

We evaluate \SYSNAME on a dataset of 18.3k popular websites, finding that, with 1M participants, we achieve 0.86 Area Under the Precision-Recall Curve (AUPRC) while providing strong (central) differential privacy guarantees ($\varepsilon = 1$).
With $\varepsilon = 10$, \SYSNAME achieves a comparable performance to a fully centralized approach (0.95 vs.~0.97 AUPRC),
and overall offers a significant improvement compared to each client only training on their local dataset (0.78 AUPRC).

Our experiments shed light on the optimal configurations balancing the practicality of deployment and detection performance.  
For instance, we show that we do not necessarily need extensive instrumentation of Javascript APIs or thousands of features, as done in previous work. 
In fact, a small set of 149 features is enough, %
even in the DP-FL setting.

Overall, while centralized techniques might miss websites and scripts due to bot detection techniques and user login requirements, the federated architecture of \SYSNAME captures real-world browsing behavior and can detect fingerprinting more robustly while providing rigorous privacy guarantees.

\section{Background}
\label{sec:bg}
This section reviews background topics, namely, Federated Learning (FL) and Differential Privacy (DP); readers familiar with them can skip it without loss of continuity.

\subsection{Federated Learning (FL)}
Federated Learning (FL) is a decentralized learning approach where participants collaboratively train a machine learning model without sharing (possibly sensitive) training data with a central server~\cite{mcmahan2016federated}.
Instead, they train local models on their individual datasets and only share model updates.
The central server only sees and aggregates the model updates and propagates the global model to the participants.

There are different ways to instantiate FL; in this paper, we follow prior work~\cite{mcmahan2018learning,naseri2022cerberus} and use the Federated Averaging (FedAvg) algorithm~\cite{mcmahan2016federated}, which builds and iteratively updates a global model.
In each round $r$, a subset of participants ($C$) is chosen from all participants by a central server.
The server sends the aggregated global model parameters from the previous round $\theta^r_{global}$ to these participants, which initialize the local model with the global model parameters (i.e., participant $i$ initializes $\theta^r_i = \theta^r_{global}$).
Each participant $i$ performs $E$ local updates using a given optimization algorithm (typically Stochastic Gradient Descent) and returns the parameters of the resulting model, $\theta^{r+1}_i$, to the server.
Finally, the server averages the local models to get $\theta^{r+1}_{global} = \frac{1}{|C|} \sum_{i \in C} \theta^{r+1}_i$.

\subsection{Differential Privacy (DP)}
DP is the established framework to define algorithms resilient to adversarial inferences.
It provides an unconditional upper bound on the privacy loss of individual data subjects from the output of an algorithm by introducing statistical noise~\cite{dwork2014algorithmic}.

\begin{definition}[Differential Privacy (DP)]
  A randomized mechanism $\mathcal{M} : \mathcal{D} \rightarrow \mathcal{R}$ is $(\varepsilon, \delta)$-differentially private if for any two neighboring datasets $d, d' \in \mathcal{D}$ and $S \subseteq \mathcal{R}$
  \begin{equation*}
    \Pr[\mathcal{M}(d) \in S]  \leq e^\varepsilon \Pr[\mathcal{M}(d') \in S] + \delta \vspace{-0.2cm}
  \end{equation*}
\end{definition}
The above definition leaves the definition of \emph{neighboring datasets} to possibly depend on the setting, and thus it can vary~\cite{mcmahan2018learning}.
The $\varepsilon$ parameter (aka privacy budget) is a numerical value ranging from $0$ to $\infty$ (lower values imply better privacy), representing the privacy loss due to the mechanism.
The additional parameter $\delta$ is referred to as the ``failure probability,'' i.e., the probability with which the mechanism fails to provide any privacy guarantees.
Therefore, $\delta$ is set to be an asymptotically small number ($\approx 10^{-5}$).

\subsection{DP in FL}
\label{subsec:dp_in_fl}

\begin{algorithm}[t]
  \caption{DP-FedAvg}
  \label{alg:cdp_fl}
  \begin{algorithmic}[1] %
      \Function{Main}{\emph{initial model} $\theta_0$, \emph{\# rounds} $R$,
        \newline \emph{\# participants} $W$, \emph{sampling probability} $q$, \emph{noise scale} $z$,
        \newline \emph{clipping parameter} $S$, \emph{optimizer} $\textrm{OPT}$}
          \State $\theta^1_{global} \gets \theta_0$
          \State $\sigma \gets \frac{zS}{qW}$
          \For{round $r$ from 1 to $R$}
            \State $P_r \gets $ randomly select participants with probability $q$

            \For{participant $k \in P_r$}
              \State $\Delta^{r+1}_k \gets \textrm{LOCAL\_UPDATE}(k, \theta^r_{global}, S, \textrm{OPT})$
            \EndFor

            \State $\theta^{r + 1}_{global} \gets \theta^{r}_{global} + \frac{1}{qW}\sum_{k \in P_r} \Delta^{r+1}_k + \mathcal{N}(0, \sigma^2I)$
          \EndFor
          \State \textbf{return} $\theta^{R+1}_{global}$
      \EndFunction
      \newline
      \Function{Local\_Update}{$k, \theta^r_{global}, S, \textrm{OPT}$}
          \State $\theta \gets \theta^r_{global}$
          \For{local epoch $i$ from $1$ to $E$}
            \State $\theta \gets \textrm{OPT}(\theta, \mathcal{D}_k)$\Comment{local update with OPT}
            \State $\Delta \gets \theta - \theta^r_{global}$
            \State $\theta \gets \theta^r_{global} + \min(1, \frac{S}{||\Delta||_2})\cdot\Delta$
          \EndFor
          \State \textbf{return} $\theta - \theta^r_{global}$\Comment{already clipped}
      \EndFunction
  \end{algorithmic}
\end{algorithm}
 
In the context of FL, DP can be defined and applied in various ways, also depending on the trust assumptions in place.

\descr{Record-level vs.~Participant-level DP.}
In FL, DP guarantees can hold at either record or participant level, depending on the definition of \emph{neighboring datasets}.
When each user contributes one record (aka a sample or a row), the dataset is defined as a collection of records.
Therefore, neighboring datasets either add or remove a single record corresponding to whether a single user contributed their record to that dataset.
Here, record-level guarantees~\cite{dwork2014algorithmic} ensure the difficulty (bound by the privacy parameter $\varepsilon$) for an adversary to determine if a single record was included in the data analysis.
Naturally, there are settings where each user contributes multiple records; e.g., in healthcare settings, each patient contributes multiple records corresponding to each hospital visit.
Similarly, in FL, each user contributes all the records in their local dataset; here, participant-level guarantees~\cite{mcmahan2018learning} are necessary.
Thus, neighboring datasets either add or remove all records that belong to a single user. 
In this case, participant-level DP ensures it is difficult (up to privacy parameter $\varepsilon$) for an adversary to determine if all the records contributed by a single user were included in the data analysis. 

While in both settings, the DP guarantees correspond to whether a single user participated in the data analysis, the resulting definitions of datasets/neighboring datasets vary due to the amount of data contributed by each user.

\descr{Central, Local, and Distributed DP.} Another difference in how DP can be integrated into FL is based on the trust placed on the server: 1) in Local DP (LDP)~\cite{pihur2018differentially}, each participant adds noise before sending updates to the server; 2) in Central DP (CDP)~\cite{mcmahan2018learning, geyer2017differentially}, participants send updates without noise, and the server applies a differentially private aggregation algorithm.

LDP has the advantage that each participant's model update is $(\varepsilon, \delta)$-differentially private; thus, not even the server is able to make inferences on them (with probability bounded by $\varepsilon$).
At the same time, however, this also means that a large amount of local noise may be required, which could severely affect utility~\cite{kairouz2021advances}.
By contrast, in CDP, individual model updates are sent unperturbed to the server; the DP guarantees are primarily with respect to the aggregate model vis-\`a-vis the other participants. 
The server is trusted with the aggregated model coefficients but not the (sensitive) training data.
Since noise is only added after aggregation, less noise is typically required, thus resulting in better model utility.

A possible alternative is to combine CDP with {\em secure aggregation} protocols.
This setting, known as Distributed DP~\cite{kairouz2021distributed}, uses CDP in that the total amount of noise added to the model updates is the same as with CDP.
Each participant also adds a \emph{small} amount of noise to their local model updates and encrypts them, using additively homomorphic encryption or secure multiparty computation, \emph{before} sending them to the server.
The server can then only decrypt the aggregate, but not the individual users' model updates.
Alas, in practice, the implementation of secure aggregation protocols in production systems is not trivial, as the details of finite precision and modular summation arithmetic are often overlooked~\cite{kairouz2021distributed}, noise has to be sampled from special discrete distributions~\cite{agarwal2021skellam,kairouz2021distributed,hartmann2023distributed}, and the computational complexity of secure aggregation protocols tend to scale with the total number of participants (typically very large in FL).

Consequently, our work relies on CDP. %
Specifically, we follow prior work~\cite{naseri2022cerberus,mcmahan2018learning} and use Algorithm~\ref{alg:cdp_fl}, which guarantees that the output of the aggregation function at the server is indistinguishable (with probability bounded by $\varepsilon$) regardless of whether a given participant shared their local model updates in training (\emph{participant-level DP}). 
Although the server is trusted with the model updates and the addition of noise during aggregation, this is a significantly weaker assumption than trusting the server with the raw training data~\cite{naseri2022local}.
Real-world deployments of differentially private FL often use CDP, such as Google's next-word prediction~\cite{mcmahan2022federated}.

\section{Browser Fingerprinting}
\label{sec:browser_fp}

We now introduce browser fingerprinting and our approach to collecting samples of fingerprinting scripts in the wild.

\subsection{What is Browser Fingerprinting?}
Browser fingerprinting is a stateless {\em online tracking} technique, usually deployed through Javascript, geared to build a unique identifier tied to a user's browser. %
Typically, fingerprinting scripts collect pieces of high-entropy device information that, combined together, yield unique and stable identifiers.

An example is shown in Script~\ref{lst:sample_fp_script}: the script builds a list of fonts installed on a device by rendering text in various fonts on a Canvas element and measuring the width of the rendered text---if the font is installed, its width will be different from when the font is not installed, in which case the user's device reverts to rendering the text in some default font.
While many fonts are pre-installed on almost all devices, it is rare to have all possible fonts installed, and thus the combination of available fonts becomes virtually unique~\cite{eckersley2010unique}.

\begin{figure}[t]
  \begin{lstlisting}[basicstyle=\linespread{0.1},style=htmlcssjs,caption=An example of a fingerprinting script~\cite{iqbal2021fingerprinting}., label={lst:sample_fp_script}]
  // Canvas font fingerprinting script.
  Fonts = ["monospace" , ... , "sans-serif"];

  CanvasElem = document.createElement("canvas");
  CanvasElem.width = "100";
  CanvasElem.height = "100";
  context = CanvasElem.getContext('2d');
  FPDict = {};
  for (i = 0; i < Fonts.length; i++)
  {
    CanvasElem.font = Fonts[i];
    FPDict[Fonts[i]] = context.measureText("example").width;
  }
  \end{lstlisting}
\reduce\reduce\reduce\reduce\reduce\reduce\reduce\reduce\reduce
\end{figure}

Although specific cases of fingerprinting  -- like the one discussed above -- are well-known, there is no consensus on one specific definition of fingerprinting~\cite{iqbal2021fingerprinting}.
If anything, it is challenging to identify the \emph{intent} behind the behavior and to determine whether it is truly used for fingerprinting purposes.
For example, while the screen resolution can be a source of high-entropy information used to potentially fingerprint users, it can also be used by modern reactive websites to display content using the appropriate layout and aspect ratio.

Thus, in this work, we follow the approach of FP-Inspector~\cite{iqbal2021fingerprinting} to identify and refer to browser fingerprinting, which uses a conservative definition based on a set of well-known heuristics and signatures.
While more details are provided in Section~\ref{subsec:ground_truth}, in a nutshell, FP-Inspector~\cite{iqbal2021fingerprinting} focuses on the four most prevalent forms of fingerprinting -- namely, Canvas, Canvas Font, WebRTC, and AudioContext -- and exclude the simple collection of properties from the \texttt{Navigator} and \texttt{Screen} APIs.

Although this might potentially miss some fingerprinting scripts and techniques (i.e., false negatives), it tends to minimize the chance of wrongly flagging scripts as fingerprinting (i.e., false positives).
This is an important requirement of fingerprinting detection, as it directly affects its validity; when deployed in the wild, false positives might amount to falsely accusing organizations of doing fingerprinting and potentially not abiding by their stated privacy policies~\cite{nhs2023}.

\begin{figure*}[t]
  \centering
  \includegraphics[width=0.99\linewidth]{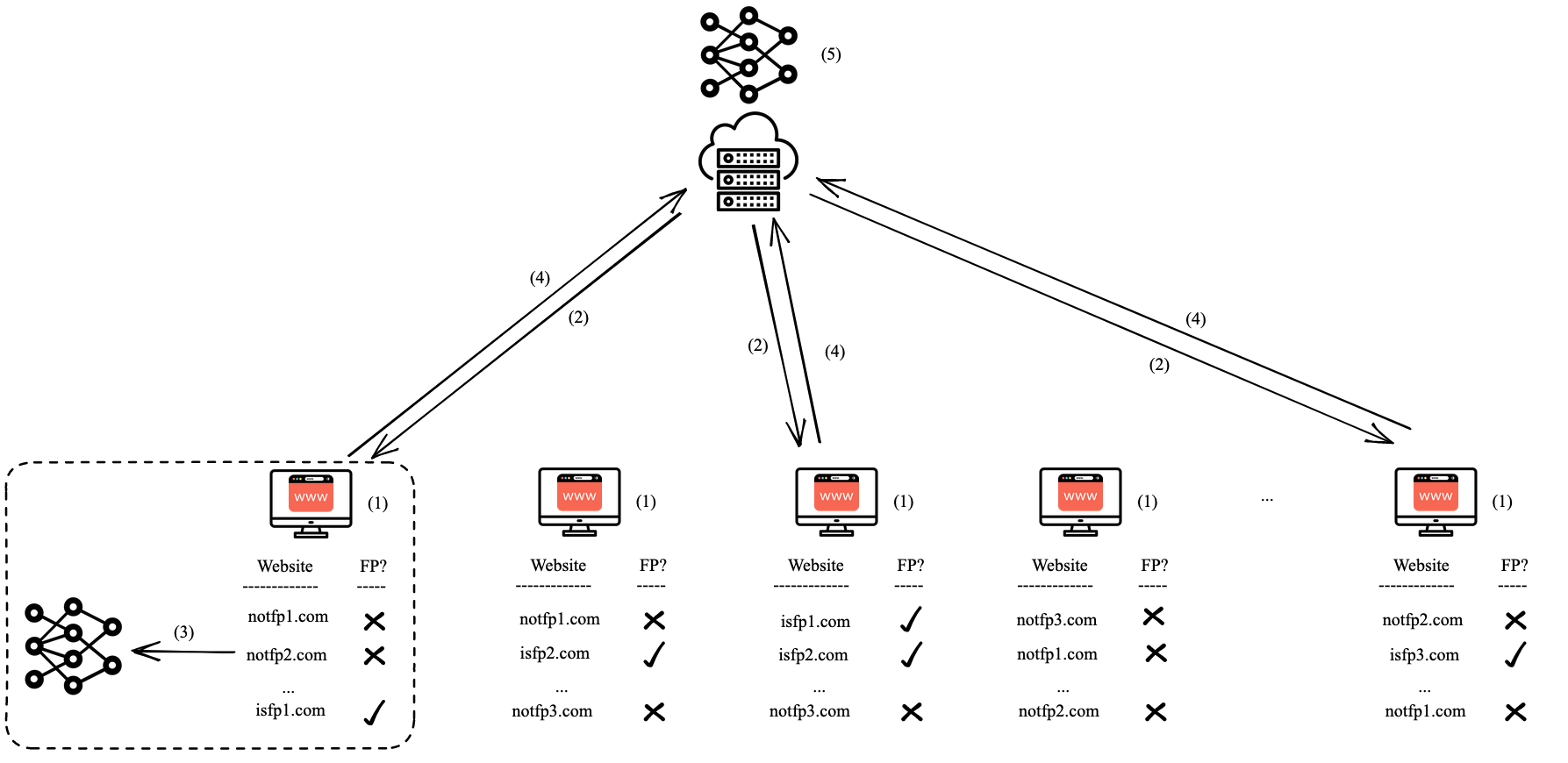}
  \caption{An overview of the \SYSNAME system: (1) Participants collect execution traces from the websites they visit. (2) At each round of training, the server selects a subset of participants and sends the previous round's global model parameters to them. (3) These participants train a local model based on data collected from websites they visited, and (4) send the local model updates to the server, (5) which aggregates them, adds differentially private noise, moves on to the next round, and repeats steps (2) to (5).}
  \label{fig:sysdiag}
\end{figure*}

\subsection{How to Collect Fingerprinting Scripts?}
\label{subsec:collect_fp_scripts}
As mentioned, fingerprinting scripts access Javascript APIs, e.g., Canvas, WebRTC, and AudioContext, to retrieve high-entropy information about the device.
Therefore, detection typically entails instrumenting browsers and collecting the content and execution traces of scripts (i.e., arguments and return values of APIs called) during web crawls.
Our experiments adhere to this approach. While our work introduces a FL approach to train ML models to detect fingerprinting in-the-wild, we collected the execution traces and page content based on a web crawl. Note that this approach, while not fully representing the distributed browsing behavior of individual users, has been adapted to simulate different browsing patterns on the web (see Section~\ref{subsec:data_collection}). %

In contrast to previous work~\cite{englehardt2016online,iqbal2021fingerprinting}, which mainly used extended versions of OpenWPM~\cite{englehardt2016online}, we use Puppeteer\footnote{\url{https://pptr.dev/}} to collect contents and traces from scripts loaded by websites.
OpenWPM uses Firefox, while Puppeteer uses Google Chrome; different browsers expose different Javascript APIs, which can, in turn, be exploited by different fingerprinting scripts. For example, although the Battery Status API was removed from Firefox since v52 (released in 2017), it is still supported in the latest version of Chrome M114.
Furthermore, as Chrome is more widely used than Firefox, we assume it is more likely that fingerprinting scripts would rely on the API surface provided by Chrome.

Moreover, rather than extracting both static (e.g., Abstract Syntax Trees) and dynamic (e.g., execution traces) features, we only focus on dynamic ones---specifically, the number of times each API is called, the arguments passed to the API, and the return values of the API call.
This follows previous work~\cite{ngan2022nowhere} showing that models built using static features are not robust, as obfuscation techniques can easily evade them.
Additionally, while Javascript APIs can be easily instrumented in the browser to collect execution traces using extensions, performing more resource-intensive operations (such as parsing Abstract Syntax Trees and running clustering algorithms) is inefficient and impractical on resource-constrained devices. 
As a result, especially in a distributed setting, it is more practical to deploy classifiers based on dynamic features alone.

\section{The \SYSNAME System}
\label{sec:system}
In this section, we introduce \SYSNAME, a privacy-preserving federated learning system to detect browser fingerprinting collaboratively.
We present an overview of the system along with its key components; then, we discuss in detail the individual steps involved in the operation of \SYSNAME.

\subsection{Overview}
Figure~\ref{fig:sysdiag} provides a high-level diagram of \SYSNAME.
The system works as follows:
\begin{enumerate}
  \item \SYSNAME participants visit websites according to their own interests and preferences. The participants run an instrumented browser (e.g., Chrome M114, which supports native API tracing) on a platform that supports FL (e.g., Android).
  \item Before federated training begins, each participant's instrumented browser performs the following actions:
  \begin{enumerate}
    \item Collects execution traces when specific (often high-entropy) monitored APIs are called;
    \item Extracts features for each script loaded on any visited website (see Section~\ref{subsec:feat_extract});
    \item Generates seed ground truth labels (fingerprinting/non-fingerprinting) for each script, according to a high-precision ground-truth heuristic (see Section~\ref{subsec:ground_truth});
    \item Participates in a pre-processing phase, sharing local summary statistics (mean and variance) of each feature with the server, which aggregates the statistics, adds DP noise, and shares them back with the participants.
    Finally, the browser scales each feature according to the global summary statistics (see Section~\ref{subsec:preproc}).
  \end{enumerate}
  \item At each round, the server selects a subset of participants to engage in that round of training.
  The server then sends the global model parameters from the previous round to these participants.
  \item Participants instantiate a local model with the global model's parameters and update their local model with the data from the websites they visited.
  \item Participants send the updated model parameters to the server.
  \item The server aggregates the updated model parameters, adds noise to satisfy DP, and generates the new global model parameters for the next round.
  \item Steps 2) to 5) are repeated over multiple rounds until the global model converges; the global model parameters are then propagated to all participants to be used for on-device browser fingerprinting detection.
\end{enumerate}

As the content served by websites and the scripts they load do not remain static over time, in theory, Steps 1) to 7) can be repeated regularly (e.g., once a month or once sufficient scripts are collected by participants) to ensure that the latest fingerprinting techniques remain detectable by the global model.
However, in our experiments, we will only consider a single data collection phase for the sake of simplicity.

As discussed in Section~\ref{sec:intro}, the data collected by our system in Steps 2a) to 2c) might reveal sensitive information that needs to be protected.
Therefore, \SYSNAME needs to provide differential privacy guarantees (satisfying CDP) when aggregating such data. That means adding statistical noise to all model updates and statistics shared with participants, including during the pre-processing phase (namely, in Steps 1d and 5).
We use DP's advanced composition theorem~\cite{kairouz2015composition} and the moments accountant~\cite{abadi2016deep} to keep track of the overall privacy budget.
In other words, \SYSNAME provides strong privacy guarantees to the participants by formally bounding leakage.

\descr{Components.} We can summarize the three main components involved in \SYSNAME as follows:

\begin{itemize}
\item {\bf\em Participants.} \SYSNAME operates in a distributed setting where participants collaboratively build a browser fingerprinting detection model based on the websites they visit.\smallskip

\item {\bf\em Server.}
The aggregation server chooses the participants for each round, propagates the global model, and aggregates the local model updates (satisfying CDP).
Although participants do not need to trust the server with their execution traces, etc., they trust it with (significantly less sensitive) model updates and that it will correctly perform differentially private aggregation.\smallskip

\item {\bf\em Model.}
While \SYSNAME can potentially be used to train any neural network-based model, in this work we instantiate a simple logistic regression model, which can be seen as a neural network with no hidden layers, one output neuron, and the sigmoid activation function.
We discuss this choice in detail in Section~\ref{sec:training}.
\end{itemize}

\subsection{Feature Extraction}
\label{subsec:feat_extract}
The features we extract from the execution traces collected by \SYSNAME's instrumented browsers include:
1) the number of times 684 potential fingerprinting Javascript APIs are called, which we refer to as the {\em API call counts},
and 2) 830 custom features. 
In total, we consider 1,514 features (684 API call counts + 830 custom features) from each execution trace.

\descr{API Call Counts.} Some APIs tap into well-known sources of high-entropy data (e.g., \texttt{Navigator.userAgent}). 
Others, such as \texttt{CanvasRenderingContext2D.measureText}, might do so only when called multiple times with a specific purpose.
These ``patterns'' can be encoded into fingerprinting ``signatures'', which can then be used in a heuristic to tag scripts as fingerprinting if they appear to match any such signatures.
FP-Inspector collects call counts from 500 APIs; however, since it uses Firefox, its crawler does not capture a number of deprecated and unimplemented APIs that are present in Chrome, e.g., \texttt{BatteryManager.level}.
Using the same instrumentation used by FP-Inspector, our Chrome-based crawler detects an additional 75 APIs accessed by scripts.

We also instrument 184 APIs available in Chrome that have been specifically flagged as ``High Entropy APIs.'' 
These are natively traced by Chrome, instead of being separately instrumented by, e.g., an extension; thus, fingerprinting scripts will not be able to evade the instrumentation.
Therefore, signals collected from those APIs are much more robust to fingerprinting evasion in JavaScript, as compared to APIs that are externally instrumented via extensions.

\descr{Custom Features.} In addition to the API call counts, we instrument and use 830 custom hand-crafted features defined by~\cite{iqbal2021fingerprinting}.
Unlike API call counts, which simply involve counting the number of times an API is called, these features are processed from the arguments and return values of the calls.
For instance, one such feature encodes whether the \texttt{WebGL2RenderingContext.fillStyle} API is set to fill the canvas with a gradient color (as opposed to a solid color).
In other words, they often represent signatures that are present across many known fingerprinting scripts and thus are very useful in detecting browser fingerprinting.

Note that FP-Inspector originally extracts 2,128 such features; however, most of them could not be properly encoded when execution traces did not contain a call to the corresponding API.
Rather, we focus on the 830 features that could be encoded even when the API was not called.
In Table~\ref{tab:sample}, we report a sample of these custom features.

Furthermore, note that, even though these features process the raw execution traces heavily, information that might identify sensitive websites might still persist.
Consider, for instance, a website that contains information about a sensitive medical condition -- e.g., \url{https://www.nhs.uk/conditions/hiv-and-aids/treatment/}.
This website might load a script that accesses a unique combination of APIs that will be visible from the API call counts.
Alternatively, the script might make an API call with unique arguments, such as storing a first-party cookie with a unique string length which will be visible from the custom features.
An adversary with access to these features might then be able to determine if a user has visited the sensitive website, thus leaking sensitive information about the user (e.g., the likelihood that the user has HIV).

\begin{table}[t]
  \centering
  \setlength{\tabcolsep}{2pt}
  \begin{tabular}{ll}
  \toprule
  {\bf API} & {\bf Operation} \\
  \midrule
  \texttt{CanvasRenderingContext2D.fillStyle} & is fill gradient? \\
  \texttt{CanvasRenderingContext2D.textAlign} & returns \texttt{start}? \\
  \texttt{CanvasRenderingContext2D.textBaseline} & returns \texttt{top}? \\
  \texttt{CanvasRenderingContext2D.lineJoin} & returns \texttt{round}? \\
  \texttt{WebGLRenderingContext.getExtension} & is first argument \texttt{EXT\_blend\_minmax}? \\
  \texttt{WebGLRenderingContext.getExtension} & is first argument \texttt{WEBGL\_draw\_buffers}? \\
  \texttt{WebGLRenderingContext.getExtension} & is first argument \texttt{WEBGL\_lose\_context}? \\
  \texttt{WebGLRenderingContext.pixelStorei} & is second argument \texttt{4}? \\
  \texttt{WebGLRenderingContext.getAttribLocation} & is second argument \texttt{r5}? \\
  \texttt{WebGLRenderingContext.depthMask} & is first argument \texttt{False}? \\
  \texttt{HTMLCanvasElement.getElementsByTagName} & is first argument \texttt{script}? \\
  \texttt{Node.isConnected} & returns \texttt{False}? \\
  \texttt{Document.getElementsByTagName} & is first argument \texttt{head}? \\
  \texttt{HTMLCanvasElement.nodeName} & returns \texttt{canvas}? \\
  \texttt{AnalyserNode.channelInterpretation} & returns \texttt{suspended}? \\
  \texttt{AnalyserNode.channelCountMode} & returns \texttt{max}? \\
  \texttt{OscillatorNode.type} & returns \texttt{triangle}? \\
  \texttt{AudioContext.state} & returns \texttt{suspended}? \\
  \texttt{RTCPeerConnection.iceGatheringState} & returns \texttt{complete}? \\
  \texttt{RTCPeerConnection.signalingState} & returns \texttt{stable}? \\
  \bottomrule
  \end{tabular}
  \caption{Sample of custom features extracted from execution traces.}\label{tab:sample}
\end{table}
  
\subsection{Ground Truth}
\label{subsec:ground_truth}
Unfortunately, there are no definite and readily available labels for Javascript scripts that can be used for fingerprinting detection, and creating them manually and at scale is hard.
Therefore, we follow an approach similar to previous work~\cite{iqbal2021fingerprinting,bahrami2021}, using high-precision heuristics to generate binary labels (fingerprinting/non-fingerprinting).
While these heuristics have high precision, they are typically narrowly defined to minimize false positives; as a result, they miss fingerprinting scripts.
However, machine learning classifiers trained on high-precision heuristics are known to generalize over fingerprinting behaviors and detect previously unknown fingerprinting techniques.
In prior work~\cite{iqbal2021fingerprinting}, machine learning classifiers were able to detect 26\% more fingerprinting scripts than manually designed heuristics in the wild.

More precisely, our ground-truth labeling heuristic is taken from~\cite{iqbal2021fingerprinting}, which uses a high-precision fingerprinting definition that minimizes the false positive rate.
This does not consider simple access to device information as fingerprinting; rather, only unwarranted accesses or aggressive calls made to well-known fingerprinting APIs are considered fingerprinting.
Below, we review the four main types of fingerprinting identified in~\cite{iqbal2021fingerprinting}, along with their ``definitions.''

\descr{Canvas Fingerprinting:} The differences in font rendering across devices are exploited to build a high-entropy identifier.
Canvas fingerprinting is considered to be happening if:
\begin{compactenum}
  \item Text is written to the \texttt{canvas} element using the \texttt{fillText} or \texttt{strokeText} method;
  \item Style is applied with \texttt{fillStyle} or \texttt{strokeStyle} method;
  \item \texttt{toDataURL} is called to extract the image from the \texttt{canvas}; and
  \item \texttt{save}, \texttt{restore} or \texttt{addEventListener} methods are not called.
\end{compactenum}

\descr{Canvas Font Fingerprinting:} Scripts rely on accessing the list of fonts installed on a device.
Canvas font fingerprinting is happening if:
\begin{compactenum}
  \item \texttt{font} property of canvas element is set to more than 20 different fonts; and
  \item \texttt{measureText} is called more than 20 times.
\end{compactenum}

\descr{WebRTC Fingerprinting:} Scripts use the access to candidate IP addresses of peers used by the WebRTC protocol~\cite{webrtc}.
The following actions define WebRTC fingerprinting:
\begin{compactenum}
  \item \texttt{createDataChannel} or \texttt{createOffer} method is called on a WebRTC peer connection; and
  \item \texttt{onicecandidate} or \texttt{localDescription} method is called.
\end{compactenum}

\descr{AudioContext Fingerprinting:} Scripts use differences in how audio signals are processed by different hardware.
The following actions define AudioContext fingerprinting:
\begin{compactenum}
  \item Either \texttt{createOscillator}, \texttt{createDynamicsCompressor}, \texttt{destination}, \texttt{startRendering}, or \texttt{oncomplete} is called.
\end{compactenum}

\subsection{Differentially Private Federated Pre-Processing}
\label{subsec:preproc}

\begin{algorithm}[t]
  \caption{DP-FedNorm}
  \label{alg:dpfed_norm}
  \begin{algorithmic}[1] %
      \Function{Main}{\emph{\# features} $F$, \emph{\# participants} $W$, 
        \newline \emph{sampling probability} $q$, \emph{noise scale} $z$,
        \newline \emph{mean clipping parameter} $S_\mu$,
        \newline \emph{var clipping parameter} $S_s$}
          \State $\sigma_\mu \gets \frac{zS_\mu}{qW}$
          \State $\sigma_s \gets \frac{zS_s}{qW}$
          \For{feature $f$ from 1 to $F$}
            \State $P^f_\mu \gets $ randomly select participants with prob.~$q$

            \For{participant $k \in P^f_\mu$}
              \State $\mu^f_k \gets \textrm{LOCAL\_MEAN}(k, f, S_\mu)$
            \EndFor
            \State $\pmb{\mu}_f \gets \frac{1}{qW} \sum_{k \in P^f_\mu} \mu^f_k + \mathcal{N}(0, \sigma_\mu^2I)$
            \newline

            \State $P^f_s \gets $ randomly select participants with prob.~$q$

            \For{participant $k \in P^f_s$}
              \State $s^f_k \gets \textrm{LOCAL\_VAR}(k, f, S_s)$
            \EndFor
            \State $\mathbf{s}_f \gets \frac{1}{qW} \sum_{k \in P^f_s} s^f_k + \mathcal{N}(0, \sigma_s^2I)$
          \EndFor
          \State \textbf{return} $\pmb{\mu}, \mathbf{s}$
      \EndFunction
      \\
      \Function{Local\_Mean}{$k, f, S_\mu$}
          \State $n \gets |\mathcal{D}_k|$\Comment{number of rows}
          \State $\mathbf{v}^f \gets f^{th}$ column from $\mathcal{D}_k$
          \State $\mu = \frac{1}{n} \sum_{i = 0}^n \mathbf{v}^f_i$
          \State \textbf{return} $\min(\mu, S_\mu)$\Comment{clip}
      \EndFunction
      \\
      \Function{Local\_Var}{$k, f, S_s$}
          \State $n \gets |\mathcal{D}_k|$\Comment{number of rows}
          \State $\mathbf{v}^f \gets f^{th}$ column from $\mathcal{D}_k$
          \State $\mu = \frac{1}{n} \sum_{i = 0}^n \mathbf{v}^f_i$
          \State $s = \frac{1}{n - 1} \sum_{i = 0}^n (\mathbf{v}_i - \mu)^2$
          \State \textbf{return} $\min(s, S_s)$\Comment{clip}
      \EndFunction
  \end{algorithmic}
\end{algorithm}

Before training, we need to normalize the extracted features to have a mean of 0 and a variance of 1.
This is known to improve model convergence.
In a centralized setting, the mean and variance of each feature can be trivially calculated by the data holder.
Whereas in a federated setting, the mean and variance of each feature have to be aggregated in a distributed way from many participants, while satisfying differential privacy.
To this end, we use Algorithm~\ref{alg:dpfed_norm}.

For each feature, $f$, after the mean ($\pmb{\mu}_f$) and variance ($\mathbf{s}_f$) are calculated, each participant scales the vector of features extracted from each script ($\mathbf{v}^f$) by computing $\frac{\mathbf{v}^f - \pmb{\mu}_f}{\mathbf{s}_f}$.
This makes the (normalized) features have a mean of 0 and a variance of 1 throughout.

\subsection{Training}\label{sec:training}
Previous work~\cite{ikram2017towards,iqbal2021fingerprinting} in the centralized setting used Support Vector Machines (SVM) and Decision Trees. %
However,~\cite{ikram2017towards} focuses on the broader problem of ``web tracking,'' building a balanced dataset where 57\% of the scripts are tracking.
By contrast, browser fingerprinting is a much more narrow definition: typically, less than 1\% of all scripts are fingerprinting.
Additionally, since SVMs do not perform well in problems with severe class imbalance~\cite{batuwita2013class}, they are not a good fit for the problem of browser fingerprinting.

Moreover, decision trees have only been recently adapted to the federated setting.
More precisely, Maddock et al.~\cite{maddock2022federated} recently introduced a framework to deconstruct the decision tree algorithm into components, with a number of settings that need to be fine-tuned to the specific application, e.g., discretizing continuous features, batching weight updates, etc.

Therefore, %
as adapting or designing novel DP-FL algorithms is not the focus of our work, we opt for a simple model architecture based on a logistic regression function, although \SYSNAME can be instantiated with any algorithm.
To ensure fast convergence, we also use the LBFGS optimizer instead of standard first-order gradient descent techniques, such as Stochastic Gradient Descent.
To the best of our knowledge, we are the first to empirically show that second-order methods such as LBFGS can be used to train models with differential privacy successfully in the federated setting.

\section{Experimental Evaluation}
\label{sec:exp_eval}
In this section, we discuss the setup of our experimental evaluation to shed light on the key factors impacting the performance of \SYSNAME.
We present our strategy to visit popular webpages and collect fingerprinting scripts.
We then discuss how we distribute these scripts among \SYSNAME participants.
Finally, we introduce the different subsets of features we experiment with.

\subsection{Dataset}
\label{subsec:data_collection}
Our first step is to collect execution traces. %
We opt to {\em simulate} a distributed setting by performing a single crawl, and then split the traces across different participants using different distributions.
We do so for two reasons: 1) arguably, it would be exceedingly difficult to recruit and instrument large numbers of users and browsers, e.g., due to privacy, efficiency, cost, and coverage issues, and 2) simulating a distributed setting enables us to experiment with script and feature distributions to evaluate the impact of different distributions on the performance of the overall model (see Section~\ref{subsec:feat_sets}). 

\descr{Popular websites.} 
Our crawl strategy follows that of prior work~\cite{iqbal2021fingerprinting}, as we visit the homepages of 20k popular websites.
We visit the top 10k sites from the Chrome User Experience Report (CrUX)~\cite{crux} and randomly sample another 10k sites ranking between 10k and 100k.
Previous studies have often used the Alexa ranking to sample and visit popular websites; we use the CrUX ranking instead since the Alexa ranking has become deprecated since May 2022~\cite{alexa2022}.
Moreover, recent research~\cite{ruth2022toppling} suggests that the CrUX dataset provides a more accurate ranking than the alternatives.

\descr{Samples.} Out of the 20k websites, we successfully visit 18,300 (91.5\%).
That is, our crawl fails to collect traces from 1,700 websites, with the overwhelming majority (64.3\%) of them due to \emph{HTTP 403 Forbidden} errors.
This error is typically returned by websites requiring user login or bot detection scripts.
Of the 18,300 websites loaded, we collect 181,633 unique scripts extracting 1,514 features each.

According to our high-precision ground-truth heuristic, 752 out of the 181,633 scripts (0.41\%) are fingerprinting.
We then split the dataset into 80\% training (145,307 scripts, 602 fingerprinting) and 20\% testing (36,326 scripts, 150 fingerprinting).

\begin{figure}[t]
  \centering
	\includegraphics[width=0.99\linewidth]{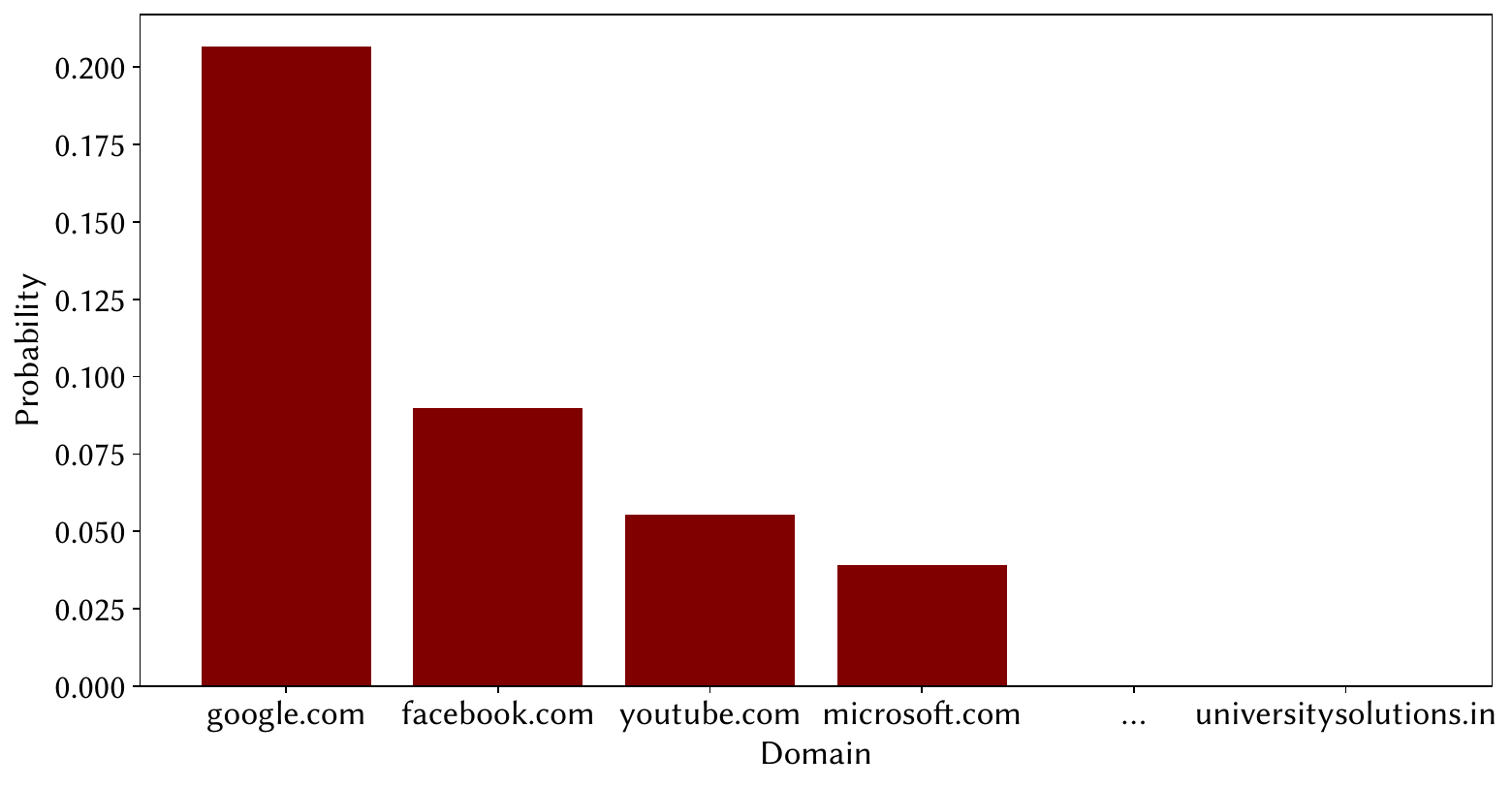}  
  \caption{Distribution of domains among participants}
  \label{fig:zipf}
  \reduce
\end{figure}

\descr{Data Distribution.}
Next, we assign the training data to the participants in \SYSNAME to simulate a realistic distributed browsing scenario. 
To do so, we build a distribution among domains based on the Tranco~\cite{victor2019tranco} list as showed in  Figure~\ref{fig:zipf}.

Although the CrUX dataset used for the crawl does provide rankings for websites, these are coarse-grained (1k, 10k, 100k, etc.). %
Therefore, we assign the fine-grained Tranco ranks to the websites present on the CrUX list.
The distribution follows the Zipf's law, which reflects the real-world observation that some websites are visited much more frequently than others~\cite{adamic2002zipf,clauset2009power}.
Thus, each participant $k$ samples a fixed set of $D$ URLs from the distribution and constructs her local list $\mathcal{D}_k$ of URLs, for which it stores all scripts that were loaded on each of these URLs from the training dataset.

\subsection{Feature Sets}
\label{subsec:feat_sets}
One key consideration of deploying FL in production is the limited computing capability available to participants~\cite{kairouz2021advances}.
While this is usually a key factor in deciding the size and complexity of models and training algorithms involved, in the context of browser fingerprinting, this consideration additionally impacts the features that the browser can feasibly extract.
As described earlier, \SYSNAME extracts two types of features: 1) API call counts and 2) custom features.
The former simply involves attaching a counter to potential fingerprinting APIs, while the latter involves reading and processing arguments and return values of APIs, which might be computationally intensive and potentially raise privacy concerns.

As a result, we experiment with different smaller subsets of the 1,514 total features, which are more practical to deploy, and investigate their impact on model performance.
Next, we define the four main feature sets we experiment with.

\descrit{All.}
This feature set comprises all 1,514 features: 1) the 500 features collected in our crawl by instrumenting the full API surface instrumented by FP-Inspector~\cite{iqbal2021fingerprinting}, along with the additional 184 APIs not included in their work but currently present in Google Chrome, and 2) 830 custom features.

\descrit{FP Inspector.} This set includes 1,330 features: 1) the 500 API call counts and 2) the 830 custom hand-crafted features.
This is the set of features considered by FP-Inspector~\cite{iqbal2021fingerprinting}. %

\descrit{JShelter.} This set consists of 588 features: 96 API call counts and 492 custom features corresponding to the APIs instrumented by JShelter~\cite{jshelter}, a browser extension that, among other use cases, includes a fingerprinting detector module.
We extract the API surface instrumented by JShelter from its source code using jsrestrictor.\footnote{\url{https://github.com/polcak/jsrestrictor/blob/master/common/fp_config/wrappers-lvl_0_1.json}}
As obtaining the exact list of custom features actually extracted by JShelter would require significant reverse engineering, we use all custom features corresponding to the API surface. %
While this may not be the exact feature set used by the extension to classify fingerprinting scripts, it provides a useful upper bound on the number of features real-world fingerprinting detectors typically have.

\descrit{High Entropy.}
This set includes 109 features: no custom features and the 109 API call counts flagged as ``high entropy'' by Chromium.\footnote{\url{https://github.com/chromium/chromium/blob/aae7191b27cef1f097b23e7742afb4895ec6a9d3/docs/privacy_budget/privacy_budget_instrumentation.md?plain=1\#L196}}

\subsection{Metrics}
To quantify the model's performance, we use the Area Under the Precision-Recall Curve (AUPRC) statistic.
Rather than choosing a single threshold for classification and calculating precision and recall for that threshold, the AUPRC summarizes the model performance across various thresholds.

Thus, AUPRC is a more robust statistic that captures the effectiveness of the model as a whole.
Also, we repeat and average all experiments over five runs.

\section{Results}
\label{sec:results}
We now present the results of our performance evaluation of \SYSNAME across various settings. 
We start with assessing the impact of federated training  -- specifically, the number of participants -- on model performance, without any differential privacy guarantees.
We then measure \SYSNAME's performance at various levels of privacy and for different feature sets used. 
In the process, we also introduce and experiment with an additional feature set called \HEP (\HEPFull). %
Next, we evaluate the performance improvements of using the feature normalization step. %
Moreover, we assess \SYSNAME with respect to Non-IID distributions.
Finally, we conduct a small-scale user study evaluating the effectiveness and computational overhead of performing a manual crawl using a prototype browser extension that deploys \SYSNAME.

\subsection{Non-DP Federated Training}

\begin{figure}[t]
  \centering
  \includegraphics[width=\mywidth\linewidth]{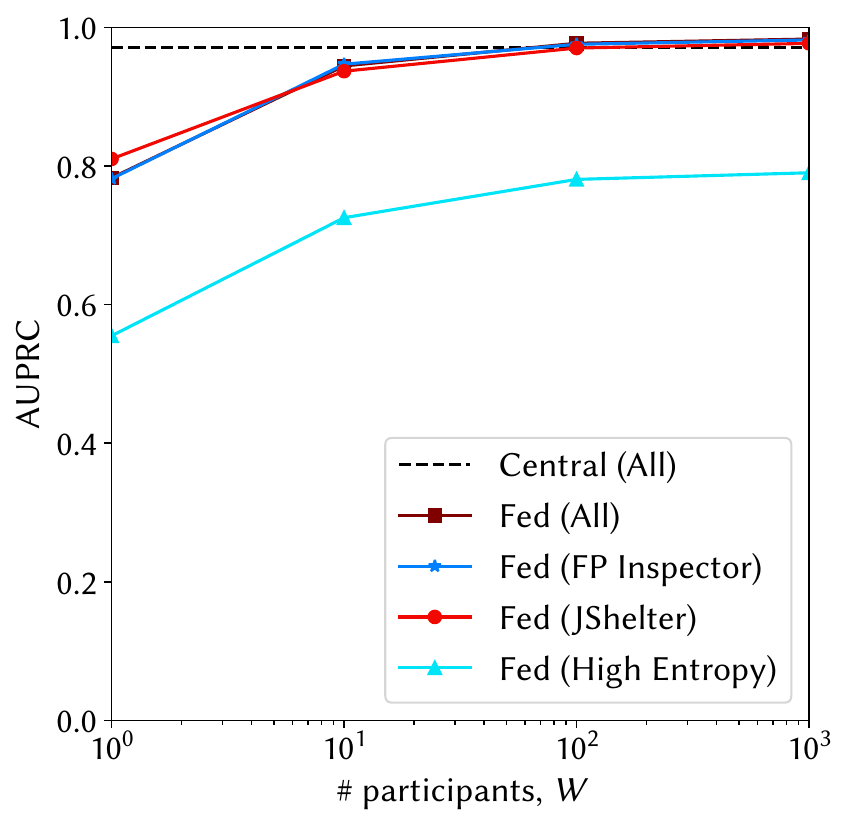}
  \caption{Model performance (AUPRC) vs. number of participants ($W$) for various feature sets.}
  \label{fig:nonprivate_per_round}
\end{figure}

We first focus on Non-DP training, where no noise is added to the feature normalization or the training algorithm ($z = 0$).
Here, we compare the impact of federated training on model performance compared to a fully centralized setting baseline.
The fully centralized setting corresponds to when a centralized crawler (e.g., FP-Inspector) can visit all websites present in our dataset.
While this baseline is the closest comparison to prior work, we emphasize that this is not necessarily the best comparison as it assumes \SYSNAME and prior work encounter the same number of FP scripts, which may not necessarily be true in practice (see Section~\ref{sec:prototype}).
Figure~\ref{fig:nonprivate_per_round} plots the model performance for an increasing number of participants with various feature sets.
In this setting, all participants train on each round ($q = 1$).

First, we observe that FL significantly improves model accuracy, across the board, compared to local training, which is the setting with $W = 1$.
Specifically, when all features are available to the classifier, AUPRC improves by 25.5\% when 1,000 participants are training with FL compared to when each participant only trains on their local dataset.

Second, regardless of the feature set, already with 100 participants, the FL model achieves optimal performance (AUPRC 0.98 when all features are available); in fact, that only marginally improves with more than 100 participants, and thus we cut our plot at $W=10^3$.
This is an important observation from a deployment perspective, as the number of participants available at each round of training is typically much lower than the overall number of participants.

Third, models trained with \textit{All}, \textit{FP Inspector}, and \textit{JShelter} feature sets all perform close to the centralized baseline, which is trained on all the scripts in the training dataset.
This suggests that having access to the additional API call counts for the APIs present in Google Chrome (\textit{All}) does not remarkably improve model performance compared to only using the APIs considered by FP-Inspector~\cite{iqbal2021fingerprinting} (\textit{FP Inspector}).
However, this might be a limitation of the high-precision ground truth heuristic used in this work, which specifically does not include certain types of fingerprinting, e.g., Battery API (we discuss this further in Section~\ref{sec:limitations}).
These types of fingerprinting will, by definition, be missed by the \textit{FP Inspector} feature set. 

Finally, even in the Non-DP setting, detecting fingerprinting based on only the API call counts of APIs natively traced by Google Chrome (\textit{High Entropy}) is not particularly effective.
Even when the number of participants increases ($W = 10^3$), with this feature set, AUPRC does not go over 0.8.

Overall, this first set of experiments indicates that FL based on logistic regression shows great promise with respect to browser fingerprinting detection -- even with a relatively small number of participants, detection performance matches that of centralized learning.

\subsection{Impact of the Privacy Parameter (\texorpdfstring{$\varepsilon$}{})}

\begin{figure}[t]
  \centering
  \includegraphics[width=\mywidth\linewidth]{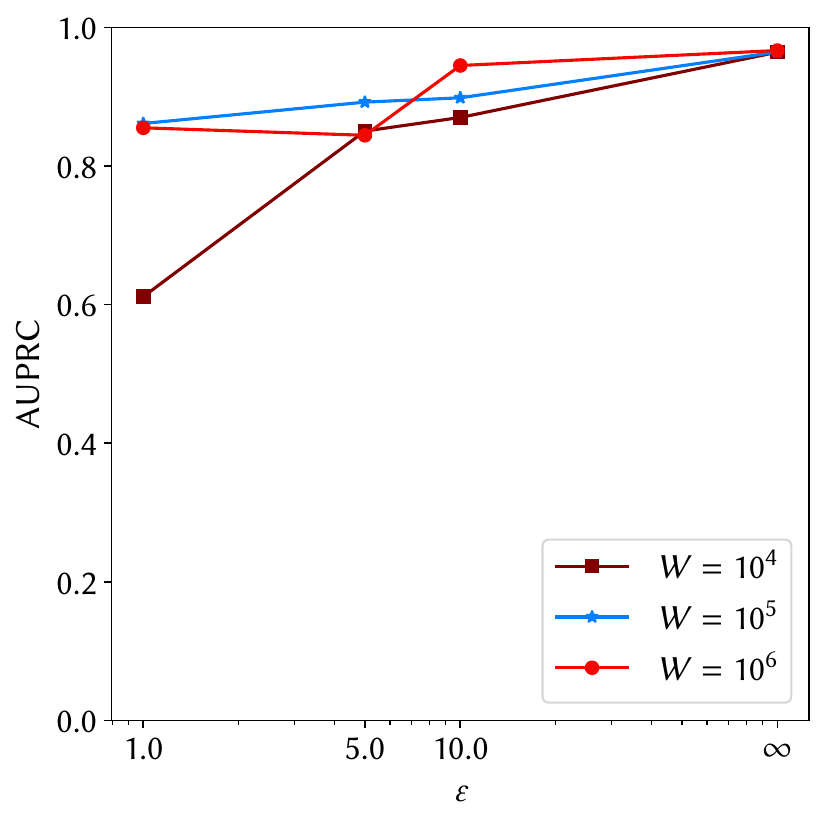}
  \caption{Model performance (AUPRC) vs. the privacy parameter ($\varepsilon$) for an increasing number of participants ($W$).}
  \label{fig:dpfed_standard_clients}
\end{figure}

We now evaluate the performance of \SYSNAME when adding differentially private noise.
We experiment with three different levels of privacy ($\varepsilon \in \{1.0, 5.0, 10.0\}$), which we refer to, respectively, as high, moderate, and low privacy.
We also include the Non-DP results ($\varepsilon = \infty$) for context.
In each round, approximately 100 participants are chosen, using Poisson random sampling, from the large pool of participants ($q = 100 / W$). 
We assume all features are available to the classifier.
Figure~\ref{fig:dpfed_standard_clients} reports model performance with various levels of privacy and varying numbers of participants.

With a small number of participants ($W = 10^4$), \SYSNAME only achieves AUPRC below 0.7, especially, at high levels of privacy ($\varepsilon = 1.0$).
However, when the number of participants increases to $W = 10^5$, the model starts to perform better, with AUPRC above 0.8 even at high privacy levels ($\varepsilon = 1.0$).
When the number of participants is high ($W = 10^6$), the model performance without DP can be recovered albeit only at a low level of privacy ($\varepsilon = 10.0$).
Additionally, when the number of participants is large ($W \geq 10^5$), the model performs similarly (within 2 standard deviations) at high and moderate privacy levels ($\varepsilon = \{1.0, 5.0\}$).
Note that this is not the number of participants per round but the total number of participants available.
This confirms that, as it is common in DP-FL, having more than a handful of participants is imperative to improve model performance, even if not all of them participate in the training each round due to privacy amplification by sampling.
In other words, for the same privacy level, the more participants exist, the less noise needs to be added.

Overall, we show that \SYSNAME achieves acceptable performance (AUPRC $\geq$ 0.8) even with high privacy guarantees ($\varepsilon = 1.0$) when enough participants are present ($W \geq 10^5$).

\subsection{Impact of the Feature Sets}
\label{subsec:feat_impact}

\begin{figure}[t]
  \centering
  \includegraphics[width=\mywidth\linewidth]{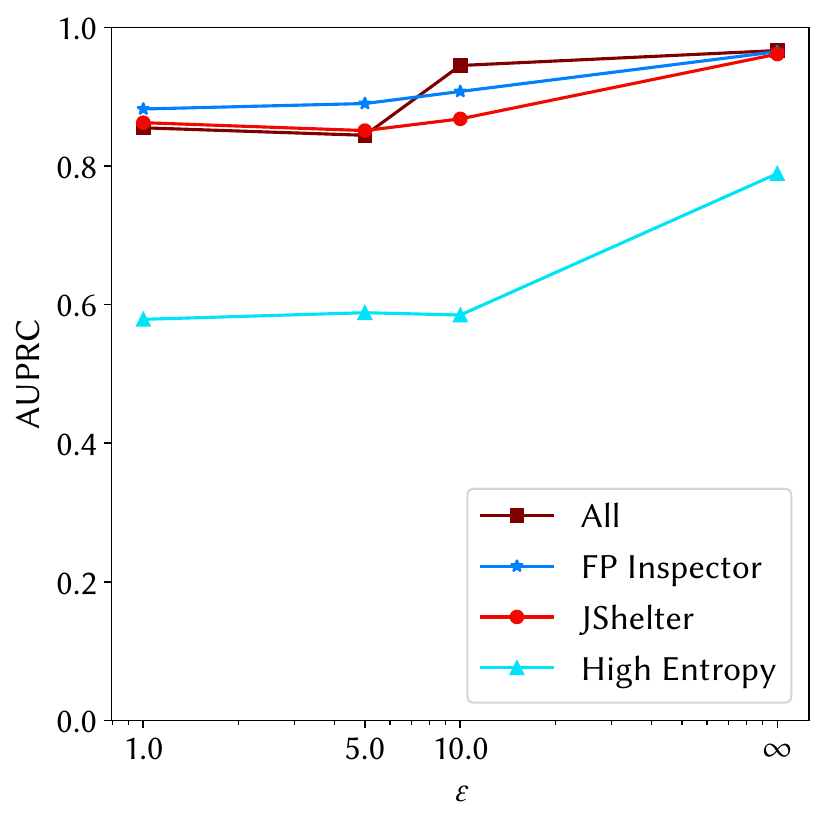}
  \caption{Model performance (AUPRC) with different feature sets for increasing levels of privacy.}
  \label{fig:dpfed_standard_feats}
\end{figure}

Next, we evaluate the feasibility of training lighter-weight classifiers with DP.
Once again, $\approx 100$ participants are chosen for each round, using Poisson random sampling.
However, the total number of participants is now fixed at $W = 10^6$.
In Figure~\ref{fig:dpfed_standard_feats}, we plot the model performance at various levels of privacy for different feature sets.

We observe that when enough features are available (\textit{All}, \textit{FP Inspector}, and \textit{JShelter}), the model performs well even at high privacy levels.
Interestingly, at moderate to high privacy levels, the \textit{JShelter} feature set has comparable performance (within 2 standard deviations) to the \textit{FP Inspector} and \textit{All} feature sets even though it only contains 40\% of the features present in the \textit{All} feature set.
This is most likely due to the differentially private feature normalization step used in \SYSNAME.
Since the noise added to the mean and variance of each feature scales with the number of features used by the classifier, having fewer but more informative features can be enough to recover the model performance compared to having more, potentially uninformative features.

Also, the model continues to perform poorly when limited to the \textit{High Entropy} feature set under differentially private training.
In fact, even at a low level of privacy ($\varepsilon = 10.0$), the model performance is very poor (AUPRC $< 0.6$), which suggests that the model has little to no utility at any reasonable privacy level.

\subsection{\HEPFull Feature Set}
One of the main objectives of \SYSNAME is to build a lightweight system that does not depend on complex hand-crafted features that are 1) hard to extract on the fly while executing scripts and 2) not robust to new types of fingerprinting that might emerge.
However, our experiments thus far show that the API call counts of APIs natively instrumented by Chrome (\textit{High Entropy} feature set) do not really yield good model performance.
Therefore, it remains an open question to determine how many features \SYSNAME requires to perform well.
More precisely, we set to answer two main questions: 1) how many APIs and 2) how many custom features are needed for \SYSNAME to reliably detect fingerprinting scripts? 

First, we sort the features extracted by our crawl according to the \emph{feature importance score}, which quantifies the impact each feature has on the final classification result.
That is, features with higher scores are more informative and correspondingly are more important to include in a minimal feature set compared to features with lower scores.
Since \SYSNAME includes a feature normalization step, the weights of a trained classifier are unaffected by the scale of each feature.
Therefore, we use the absolute weights of the trained classifier as our feature importance score.
After sorting according to feature importance, we add additional features to the \textit{High Entropy} feature set (more informative features are added first) and plot the model performance against the number of features added.

\begin{figure}[t]
  \centering
  \includegraphics[width=\mywidth\linewidth]{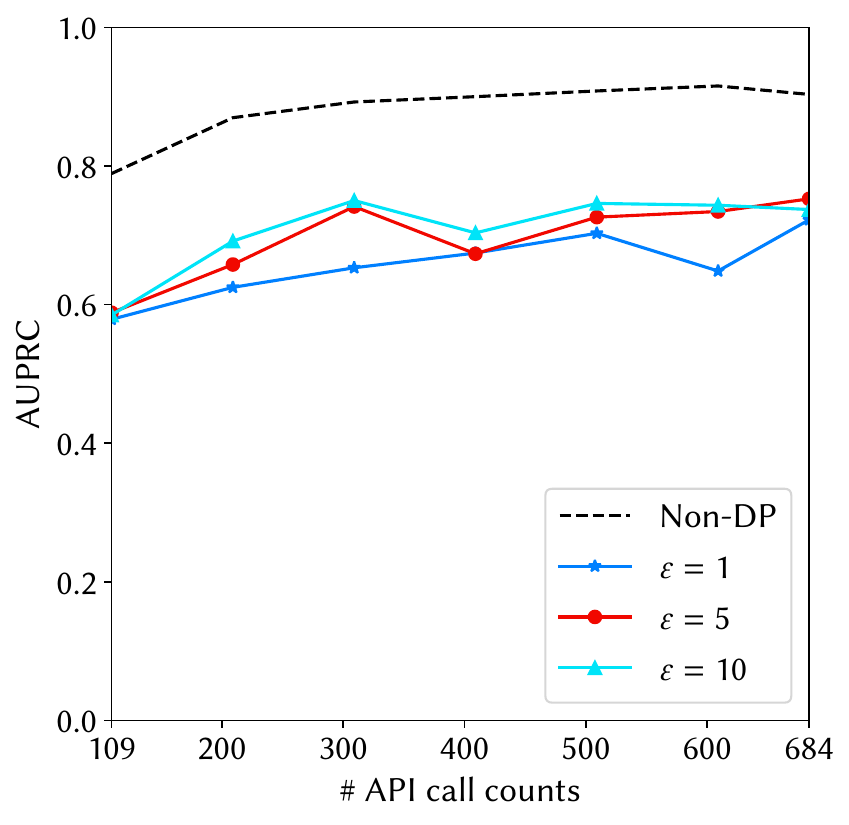}
  \caption{Model performance (AUPRC) with an increasing number of API call counts only available to the model.}
  \label{fig:high_entropy_apis_only}
\end{figure}

\descr{Adding API Call Counts.}
Figure~\ref{fig:high_entropy_apis_only} plots the model performance as the number of API call counts available to the model increases from 109 (\textit{High Entropy} feature set) to 684 (total number of API call counts captured by the crawl) at various privacy levels.
We set the total number of participants to $W = 10^6$ as before.
Regardless of the number of API call counts available to the model, the performance only improves marginally and remains low (AUPRC $< 0.8$).
In fact, even when there is no DP noise added, there is a significant drop in model performance of 6.8\% from a model with access to all features, which answers the first question: no number of APIs (for which only call counts are collected) by themselves is enough for the model to reliably detect browser fingerprinting.

\begin{figure}[t]
  \centering
  \includegraphics[width=\mywidth\linewidth]{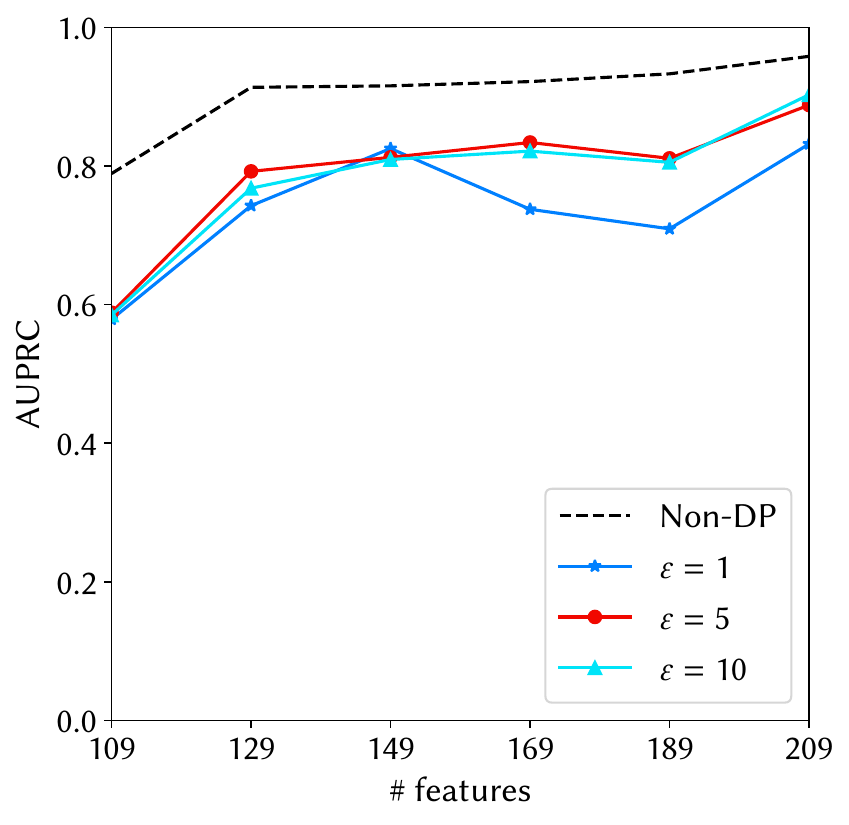}
  \caption{Model performance (AUPRC) with an increasing number of (possibly custom) features available to the model.}
  \label{fig:high_entropy_custom}
\end{figure}

\descr{Adding API Call Counts \& Custom Features.}
Figure~\ref{fig:high_entropy_custom} plots the model performance as the number of (potentially custom) features increases from 109 (\textit{High Entropy} feature set, API call counts only), in 5 small increments of 20 features, to 209. (We stop after adding 100 additional features, as we find an optimal configuration where the model performance improves significantly after adding a small number of features.)

Once again, we set $W = 10^6$.
When $\varepsilon \geq 5.0$, the performance of the model improves with the number of features available, plateauing after 20 additional features are added to the \textit{High Entropy} feature set.
For $\varepsilon = 1.0$ on the other hand, AUPRC improves until 40 additional features are added, after which performance momentarily drops.
This is most likely due to the differentially private feature normalization step, where the number of features impacts the amount of noise in training, therefore causing a tradeoff between adding more (informative) features and adding more noise in training.
In general, across all privacy levels, we find that 40 additional features provide an optimal tradeoff between adding a minimal number of features and achieving good model performance.
More precisely, these 40 features consist of 17 API call counts and 23 custom features.
This shows that to achieve good model performance, custom features are indeed necessary---although a minimal set of 23 out of the original 830 features seems to be enough.

Among these 40 additional features, we find the \texttt{BatteryManager.level} API call count to be highly informative in detecting browser fingerprinting behavior.
Note that Battery API was not used in defining the high-precision ground truth heuristic.
Rather, we find that it is often used {\em together} with other APIs used by the heuristic.

\begin{figure}[t]
  \centering
  \includegraphics[width=\mywidth\linewidth]{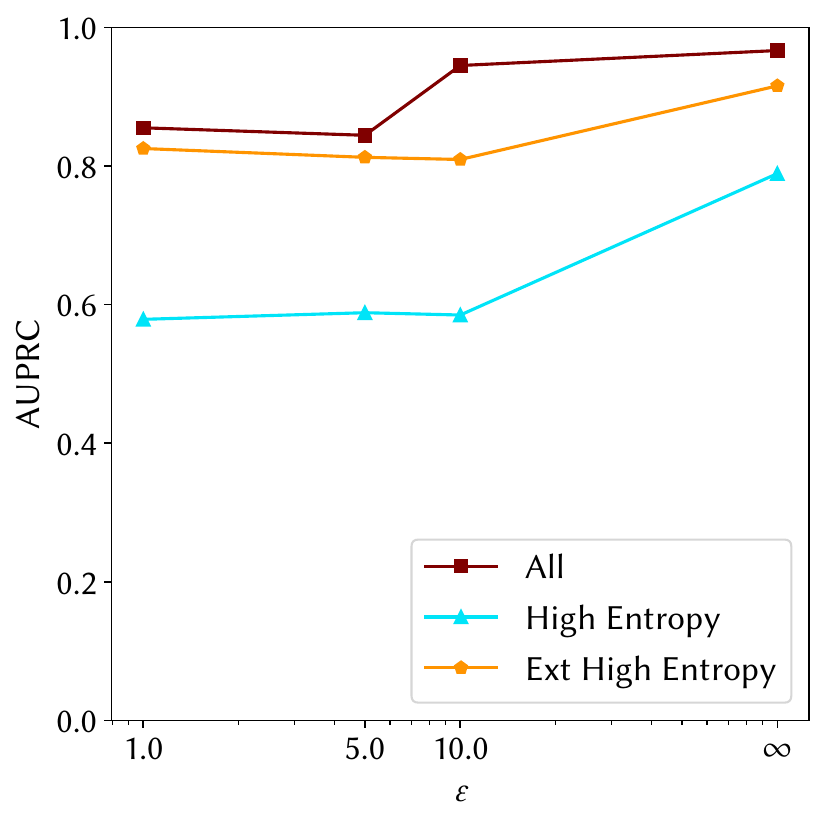}
  \caption{Model performance (AUPRC) vs $\varepsilon$ for various feature sets.}
  \label{fig:dpfed_standard_feats_high_entropy}
\end{figure}

\descr{The \HEP set.}
We define the union of the \textit{High Entropy} feature set and the 40 additional features identified above as the \HEPFull (\HEP) feature set.
In Figure~\ref{fig:dpfed_standard_feats_high_entropy}, we compare the model performance with this set to \textit{All} and \textit{High Entropy} feature sets at various levels of privacy.
As expected, \HEP yields much better performance at all levels of privacy than \textit{High Entropy}.
Furthermore, as observed with the \textit{JShelter} feature set earlier, at a high privacy level ($\varepsilon = 1.0$), \HEP performs comparably to \textit{All} (most likely due to feature selection) with only 9.8\% of the features available to the former.

\subsection{Impact of Feature Normalization}

\begin{figure}[t]
  \centering
  \includegraphics[width=\mywidth\linewidth]{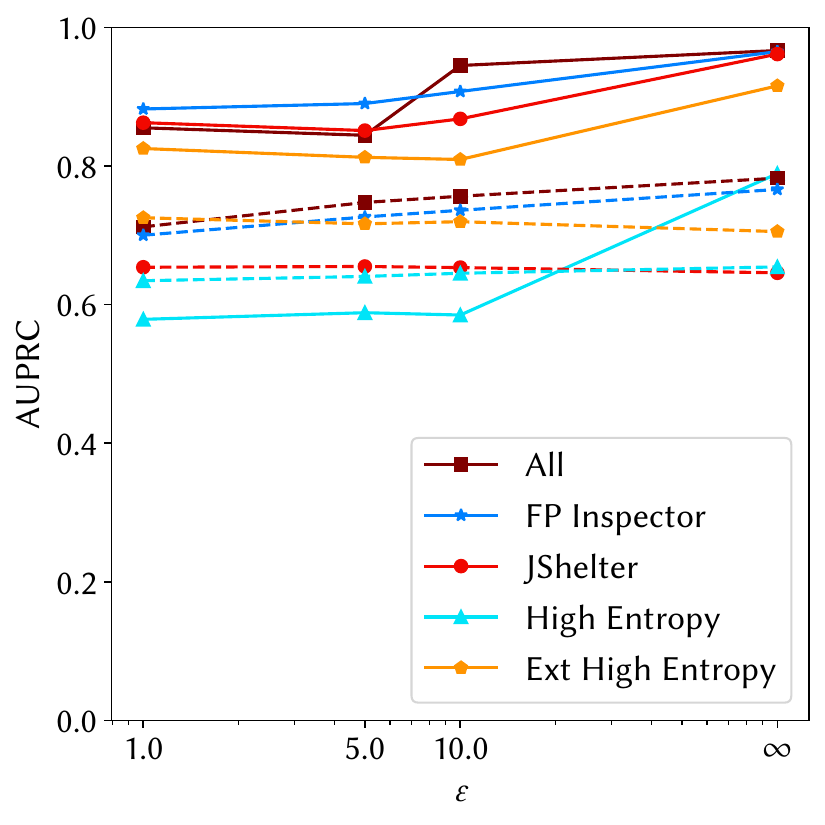}
  \caption{Model performance (AUPRC) vs $\varepsilon$ for various feature sets. The dotted lines represent results without feature normalization. %
  }
  \label{fig:feat_norm}
\end{figure}

To the best of our knowledge, ours is the first work to use differentially private feature normalization in the FL setting.
Therefore, we also investigate its effects on performance.
To that end, Figure~\ref{fig:feat_norm} plots the model performance at various levels of privacy for different feature sets.
Specifically, we plot the model performances when \SYSNAME is run with (in solid lines) and without (in dotted lines) the feature normalization step.

First, except for a few settings (\textit{High Entropy} feature set, $\varepsilon \leq 10.0$), adding the differentially private feature normalization step always improves model performance.
However, this is a specific setting where the model has access to very little, possibly uninformative features; thus, model training is easily impacted by the addition of even small amounts of noise.

When the model has access to richer custom features, it always performs appreciably better due to feature normalization.
Specifically, at a high privacy level ($\varepsilon = 1.0$), model performance improves by up to 20.8\% (\textit{JShelter} feature set).

Consequently, we believe that feature pre-processing can indeed be an important step when training with DP.
Allocating privacy budgets for feature pre-processing steps, such as normalization, can be advantageous, even though this may come at the cost of higher noise at each round of training.

\subsection{Impact of Non-IID distributions}

\begin{table}[t]
\small
\centering
\setlength{\tabcolsep}{6pt}
\begin{tabular}{rr}
\toprule
{\bf \%Participants with} & {\bf Non-IIDness} \\
{\bf Limited Knowldege} & {\bf Score}\\
\midrule
0 & 1.00 \\
50 & 6.50 \\
100 & 9.05 \\
\bottomrule
\end{tabular}
\caption{Non-IIDness scores vs. percentage of participants with limited knowledge, with 0\% corresponding to an IID distribution and 100\% to an extreme non-IID distribution.}\label{tab:non-iidness}
\reduce
\end{table}

\begin{figure}[t]
  \centering
  \includegraphics[width=\mywidth\linewidth]{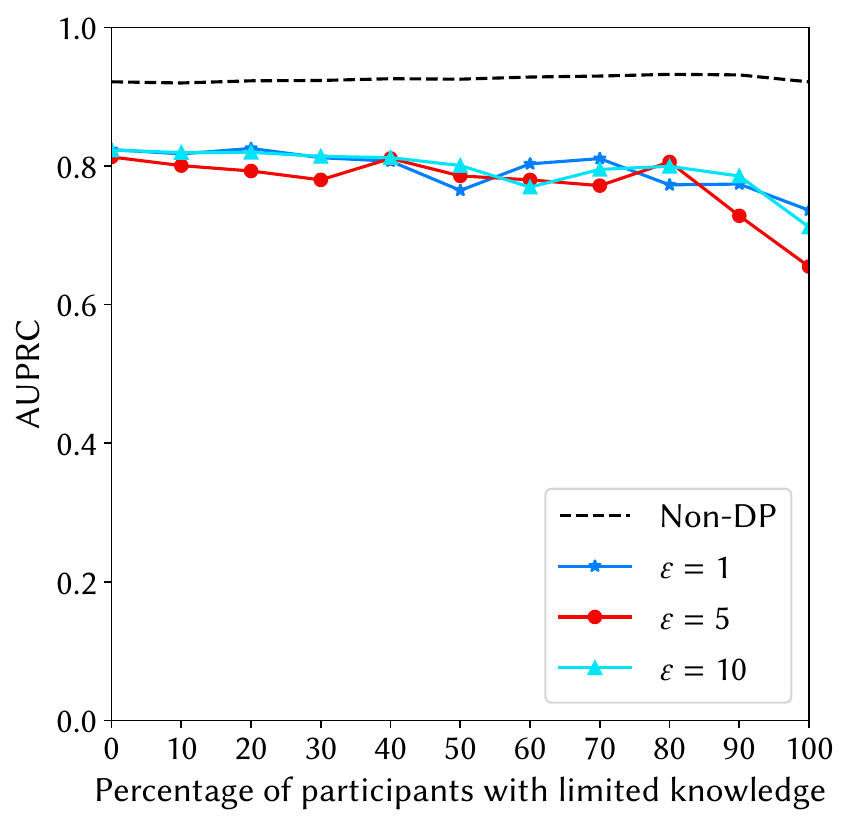}
  \caption{Model performance (AUPRC) vs. various percentages of participants with limited knowledge.}
  \label{fig:non_iid}
\end{figure}

So far, participants have drawn scripts from domains according to the Tranco ranking, thus resulting in Independent and Identically Distributed (IID) training data.
However, in practice, training data might be distributed in more complex ways, which might impact the performance of \SYSNAME.

Consequently, we stress-test \SYSNAME with respect to increasingly Non-IID distributions by introducing {\em ``Participants with Limited Knowledge''}, following recent work by Naseri et al.~\cite{naseri2022cerberus}.
These only have access to at most one type of fingerprinting script (Canvas, Canvas Font, Audio, or WebRTC), thus yielding a different distribution of fingerprinting scripts encountered by each participant.
By varying the percentage of these participants in \SYSNAME, we vary the Non-IIDness of the distributions and quantify ``how Non-IID'' the resulting distributions are using the Non-IIDness score from~\cite{naseri2022local}; please refer to Appendix~\ref{app:noniidness} for details on how the score is computed.
We report the Non-IIDness scores for three settings in Table~\ref{tab:non-iidness}, which confirms that the Non-IIDness level does increase with increasing numbers of participants with limited knowledge.

We then fix the number of participants to $W = 10^6$, use the \HEP feature set, and plot the corresponding AUPRC scores at various privacy levels in Figure~\ref{fig:non_iid}.
As expected, the model performance drops with more participants with limited knowledge.
However, the drop is pronounced only when more than 80\% of participants have limited knowledge (with $\varepsilon = 5$), which is highly unlikely in practice.
Note that, although $\varepsilon = 1$ appears to yield higher AUPRC than $\varepsilon = 5$ with 100\% limited-knowledge participants, the scores are actually within $\pm$ standard deviation of each other. 
Therefore, we conclude that \SYSNAME is reasonably resistant to Non-IID distributions which might be encountered in real-world settings.

\subsection{Prototype Deployment}
\label{sec:prototype}
To evaluate the deployability of \SYSNAME, we built a prototype Chrome extension that performs the data collection steps (Steps 2a to 2c).
These steps have the greatest impact on browser performance as every call to the Javascript APIs from every script has to be instrumented and captured, while the machine learning steps might only happen once a day.

\descr{Effectiveness.} We manually visit the top 300 domains from the Tranco ranking, excluding domains that do not expose a website (e.g., content delivery networks) and adult entertainment websites.
In this manual crawl, we imitate real user interactions by logging in using a temporary Google account whenever possible, solving CAPTCHAs, and consenting to all cookie notices.
We then compare the number of fingerprinting scripts encountered by the manual crawl with that of two automated crawlers (a basic and an advanced one).
Both automated crawlers use the same Puppeteer architecture used in the rest of the paper.
However, the advanced crawler uses the \textit{puppeteer-extra-stealth-plugin}~\cite{puppeteer} commonly used in prior work~\cite{jordan2021towards,rack2023jack} that includes bot detection evasion techniques that are continuously updated to remain highly effective.
By closely mimicking human-operated browsers~\cite{jordan2021towards}, the plugin enables the browser to appear legitimate, and crucially bypass bot detectors that might otherwise prevent a crawler from visiting a website.

\begin{table}[t]
\small
\centering
\setlength{\tabcolsep}{4pt}
\begin{tabular}{r|rrr}
\toprule
{\bf Type of FP} & {\bf Manual} & {\bf Basic Auto} & {\bf Advanced Auto} \\
\midrule
{\bf Canvas} & 189 & 49 & 40 \\
{\bf Canvas Font} & 8 & 8 & 5 \\
{\bf Audio} & 32 & 27 & 21 \\
{\bf WebRTC} & 49 & 10 & 8 \\
\midrule
{\bf Total} & 220 & 54 & 43 \\
\bottomrule
\end{tabular}
\caption{Number of fingerprinting scripts captured by manual vs. automated crawls. {\em NB:} The same script can perform multiple types of fingerprinting.}\label{tab:fp_man_vs_aut}
\end{table}

The difference in the number of fingerprinting scripts collected by each crawl technique is reported in Table~\ref{tab:fp_man_vs_aut}.
This showcases the advantage of \SYSNAME: when a real user logs in and goes through the authentication flow, the system captures 3.07$\times$ more fingerprinting scripts than the automated crawlers.
Additionally, although the advanced crawler encounters slightly fewer fingerprinting scripts than the basic counterpart, we believe this may be because bot detection scripts may only fingerprint suspicious activity.
Since the crawler deploys evasion techniques against bot detection, it may not appear suspicious enough to be fingerprinted as often as the basic crawler, thus resulting in this discrepancy.

\descr{Computational Overhead.} Finally, we compare the computational overhead of deploying \SYSNAME using the lighthouse performance metrics, which are widely used to evaluate the performance of webpages.\footnote{\url{https://github.com/GoogleChrome/lighthouse}}
The metric produces an overall ``score'' between 0 and 1, which is the weighted average of several factors, such as the time taken for the first element to be painted and for the webpage to be fully interactive.
We repeat the top-300 crawl with and without the prototype extension and evaluate the difference in the overall ``score'' assigned by the metric.
We find that there was a mean performance drop of 0.0018 and a maximum drop of 0.078 when the extension is enabled, which shows that \SYSNAME introduces a negligible computational overhead.

\section{Related Work}
\label{sec:related_work}
In this section, we review prior work on browser fingerprinting and Federated Learning (FL).

\subsection{Browser Fingerprinting}
Early research on detecting browser fingerprinting mainly resulted in the creation and manual curation of blocklists, managed by vendors such as Disconnect~\cite{disconnect2018}, EasyList~\cite{easylist}, EasyPrivacy~\cite{easyprivacy}, and PrivacyBadger~\cite{privacybadger}.
However, maintaining these lists can be difficult due to constantly changing Web APIs and fingerprinting techniques; thus, more sophisticated methods have been proposed, ones that rely on hand-crafted heuristics and machine learning. %

\descr{Heuristics-based detection.}
Acar et al.~\cite{acar2013fpdetective} and Roesner et al.~\cite{roesner2012detecting} use a combination of heuristics and manual investigations to detect fingerprinting in the wild.
In follow-up work, Acar et al.~\cite{acar2014web} introduce the first heuristic to detect canvas fingerprinting automatically, based on the arguments and return values of \texttt{fillText}, \texttt{strokeText}, and \texttt{toDataURL}.
Englehardt and Narayanan~\cite{englehardt2016online} expand on these heuristics to detect three additional types of fingerprinting: Canvas Font, WebRTC, and AudioContext.
This heuristic was also used by Iqbal et al.~\cite{iqbal2021fingerprinting} and Das et al.~\cite{das2018web}.
The former also note that these heuristics often have to be defined narrowly, thus leading to fingerprinting scripts potentially being missed.
Additionally, maintaining these heuristics in the face of a constantly evolving web platform can be difficult.

\descr{Machine Learning-based detection.}
Ikram et al.~\cite{ikram2017towards} use a one-class SVM to detect browser fingerprinting scripts using static features extracted from the Javascript code.
Iqbal et al.~\cite{iqbal2021fingerprinting} build a Decision Tree classifier that used features based on both dynamic (execution traces) and static (abstract syntax trees) analysis. %
However, as both studies trained classifiers on a centralized dataset, they might potentially miss webpages that only load fingerprinting scripts \emph{after} waiting for user interaction or after a user has logged in~\cite{acar2013fpdetective}.

\descr{Other mitigation strategies.}
Extensions like Canvas Defender~\cite{canvasdefender} and privacy-focused browsers like Tor~\cite{torbrowserfp} and Brave~\cite{bravebrowserfp} address browser fingerprinting by either normalizing Javascript APIs to return exactly the same value for all devices, or by randomizing the outputs of potential fingerprinting APIs.
However, in almost all cases, this leads to some amount of website breakage and loss in functionality~\cite{iqbal2021fingerprinting}.
For example, Tor keeps the browser window at a standard size to prevent it from being used for fingerprinting, thus disabling the ability for users to maximize it to fit their screens.

\subsection{DP in FL}
DP is being increasingly integrated into FL applications, ranging from next-word prediction~\cite{mcmahan2018learning} to medical image analysis~\cite{adnan2022federated}, and spam classification~\cite{vasyl2022}.
As discussed in Section~\ref{subsec:dp_in_fl}, the DP variants used in FL can be distinguished as Local DP (LDP) or Central DP (CDP).
While LDP has been deployed in FL applications~\cite{truex2020ldp,naseri2022local,sun2021ldp}, it often reduces model performance even at moderately large privacy parameters.
On the other hand, CDP has been more successfully applied to FL~\cite{mcmahan2018learning,naseri2022local,naseri2022cerberus,adnan2022federated} and has been deployed in large-scale production systems~\cite{kairouz2021practical}.
However, CDP requires the users to trust the aggregation server with their raw model updates and to aggregate and add differentially private noise correctly.
Relaxing this trust assumption can make inference attacks possible~\cite{nasr2019comprehensive,melis2019exploiting}.

In our work, we chose the CDP approach for FL as this is the first work applying federated learning to the problem of browser fingerprinting detection.
Providing formal DP guarantees in this setting is already challenging due to the significant class imbalance between fingerprinting and non-fingerprinting scripts.
Therefore, we focus on experimenting with and adapting DP-FL techniques rather than introducing novel DP release algorithms or FL schemes.

\subsection{FL for Security Tasks}
Previous work has also used FL in security-oriented applications, e.g., to detect or predict security events~\cite{li2020deepfed,liu2020deep,naseri2022cerberus}.
This makes sense as it is often useful to have multiple data contributors, given that these are often relatively rare events.
However, as data could be sensitive, data owners may not be willing to openly share them with a central repository.
Therefore, FL has been used to fill this gap, enabling more accurate machine learning models to be built without compromising the privacy of the data owners.

For instance, Cerberus~\cite{naseri2022cerberus} uses FL to train a recurrent neural network (RNN) model to predict future security events based on past events contributed by participating organizations.
DeepFed~\cite{li2020deepfed} is a framework that collaboratively trains a deep neural network to detect intrusions into cyber-physical systems (CPSs).
Furthermore, Liu et al.~\cite{liu2020deep} present a framework that collaboratively trains an attention-based deep neural network model to detect anomalies in Industrial Internet of Things (IoT) devices.
Additionally, FL has also been used to detect malware and intrusions into IoT and mobile devices~\cite{rey2022federated,galvez2020less,kang2019incentive,mothukuri2021federated}.

\section{Discussion \& Conclusion}
\label{sec:conclusion}

\subsection{Summary}
In this paper, we experimented with using Differentially Private Federated Learning (DP-FL) in the context of browser fingerprinting detection.
We introduced and evaluated \SYSNAME, a federated system to detect fingerprinting based on the collective browsing behavior of many users while providing formal differential privacy guarantees.
We collected fingerprinting scripts in the wild using Puppeteer, Google Chrome, and the CrUX website ranking.
This allowed us to capture more APIs than prior work and more accurately characterize the prevalence of browser fingerprinting on popular websites.

We conducted several experiments evaluating the impact of different feature sets and privacy levels, aiming to assess the impact of real-world constraints on model performance.
We also developed a model based on a minimal feature set comprising only 149 API call counts and 23 custom features, which achieves AUPRC above 0.8 even at high privacy levels ($\varepsilon = 1.0$).
This is a significantly smaller feature set compared to prior work that used 500 API call counts and 2,128 custom features and is thus far more practical for real-world deployment on end-user devices. %

\subsection{Privacy \& Robustness}
As mentioned, from a security point of view, there are two main challenges when using FL in real-world applications: privacy and robustness to poisoning attacks.

\descr{Membership/Property Inference.} Previous work~\cite{melis2019exploiting} has shown that membership and property inference attacks are possible in FL when only a handful of participants are involved (e.g., less than 100).
However, \SYSNAME is not vulnerable to them as it is meant to be deployed in settings with many users.
Moreover, it provides (Central) DP guarantees, which provably mitigate membership inference attacks, as recently shown by Naseri et al.~\cite{naseri2022local}, with $\varepsilon = 5.8$ and as little as four participants.
Furthermore, by providing participant-level instead of record-level DP guarantees, we also protect against property inference.

\descr{Malicious Servers.} Model updates are sent to the \SYSNAME server unperturbed/unencrypted; thus, a malicious server can potentially infer the presence of specific records in users' training data~\cite{nasr2019comprehensive}.
However, as discussed in Section~\ref{subsec:dp_in_fl}, in this context, we believe that trusting the server with access to model updates -- rather than raw data -- is a reasonable compromise.
Our work is the first to use FL in the context of browser fingerprinting; we focus on assessing the \emph{feasibility} of this approach and measuring the impact of different settings (data distribution, feature sets, privacy levels, number of participants, etc.), while leaving it to future work to relax this assumption by extending \SYSNAME to support Local or Distributed DP.

\descr{Poisoning Attacks.} Data poisoning and backdoor attacks are also significant concerns in FL due to the distributed nature of the computation with untrusted participants~\cite{bagdasaryan2020backdoor}.
In generic data poisoning attacks, adversarial participants attempt to compromise the utility of the global model by contributing malicious model updates.
In targeted (aka {\em backdoor}) attacks, the adversary wants the FL system to deliberately misclassify specific samples or records to an adversary-defined class.
In the context of browser fingerprinting, malicious participants could try to mount backdoor attacks so that \SYSNAME will classify fingerprinting scripts as non-fingerprinting. 
However, even though previous work~\cite{naseri2022local,naseri2022cerberus} shows that CDP can defend against backdoor attacks,
we leave it to future work to include a full experimental evaluation of data poisoning attacks against \SYSNAME.

\subsection{Practical Deployments}
An important aspect of the real-world deployability of \SYSNAME is %
whether a large range of devices with different computing power and resource constraints %
can support it.
Browser fingerprinting detection is inherently relevant to a large variety of devices, from powerful computers/laptops to mobile devices and embedded devices.
This setting is typically referred to as cross-device FL~\cite{bonawitz2019towards}.
Therefore, having a lightweight FL system is very important. %
The choice of using a simple logistic regression model rather than a deep neural network means that models are small and training them does not require specialized hardware like GPUs.

Furthermore, as discussed in Section~\ref{subsec:feat_sets}, the complexity of the features that can be extracted from the scripts is also affected by the capabilities of the devices.
In previous work, fingerprinting detectors are deployed through browser extensions like JShelter~\cite{jshelter}, instrumenting well-known APIs used for fingerprinting.
However, a relevant percentage of web browsing happens on mobile devices where browser extensions cannot always be installed.\footnote{As of May 2023, 52\% of Web traffic is from mobile, as reported from \url{https://gs.statcounter.com/platform-market-share/desktop-mobile-tablet/worldwide\#monthly-202102-202202}}
Therefore, using browser extensions as a deployment method is a significant limitation with respect to learning fingerprinting behaviors. 

By contrast, our work focuses on natively traced APIs by Google Chrome.
By doing so, \SYSNAME can be deployed directly in the browser, regardless of the kind of device, and increase the coverage of participants, thus improving utility as a whole.
Furthermore, by restricting to a small subset of features, we ensure that \SYSNAME does not add significant performance costs to normal browser operation across a wide range of devices.

\subsection{Implications for Browsers}
The web ecosystem is crucial for an accessible and free Internet. 
As such, it has to continue to bring powerful capabilities for developers and users to complement dedicated apps available on mobile and desktop platforms.
However, powerful web capabilities and APIs need to be evaluated for their potential to track end users' devices. 
Any API with a sufficiently high entropy can simultaneously be used for legitimate purposes and abused as a fingerprinting surface.
Therefore, although this is relevant not only to \SYSNAME but to the fingerprinting ecosystem as a whole, we advocate for a common way to evaluate new web APIs for ``fingerprinting potential'' before they are deployed. This type of evaluation could guide their specifications and implementation to balance the creation of new capabilities with legitimate privacy and tracking concerns for users.

For example, while the recently announced WebGPU API~\cite{webgpu} can significantly assist high-performance computations and complex graphics rendering from within the browser, it might also enable potential fingerprinting scripts to more precisely identify the GPU hardware present in a machine compared to the old WebGL standard.
In fact, this has already been acknowledged in the working draft of the WebGPU specification as increasing the risk of browser fingerprinting from the old WebGL standard~\cite{webgpuw3c}.

\subsection{Limitations \& Future Work}\label{sec:limitations}

\descr{Simulating distributed settings.} Although \SYSNAME is designed to learn from distributed real-world user behavior, due to cost and privacy challenges with collecting real-world browsing data (e.g., execution traces) from a large number of users, we chose to gather fingerprinting scripts using a centralized crawl and instead {\em simulate} a distributed setting.
Having demonstrated a proof of concept for \SYSNAME and its effectiveness, our next step is to focus on these challenges and evaluate its performance in an actual distributed setting.

\begin{table}[t]
\small
\centering
\setlength{\tabcolsep}{4pt}
\begin{tabular}{r|rrrr}
  \toprule
  $W$            &      $  \varepsilon{=}1$        & $\varepsilon{=}5$                       & $\varepsilon{=}10$                        &   No DP       \\
\midrule
$10^4$ & 0.61 & 0.85 & 0.87 & 0.96 \\
$10^5$ & 0.86 & 0.89 & 0.90 & 0.96 \\
$10^6$ & 0.86 & 0.85 & 0.94 & 0.97 \\
\bottomrule
\end{tabular}
\caption{Summary of Area Under the Precision-Recall Curve (AUPRC) values with different numbers of participants ($W$) and levels of privacy (using the \textit{All} feature seat). For comparison, a fully centralized setting yields 0.97 AUPRC, while local only results in an average of 0.78 AUPRC.}\label{tab:results}
\end{table}

\descr{Accuracy.} 
\SYSNAME achieves reasonably high accuracy even at strong privacy levels (with 1M participants, 0.86 AUPRC with $\varepsilon = 1$)
and significantly improves over the local-only setting, i.e., all clients only train on their local datasets (0.78 AUPRC).
However, there is still a non-negligible drop in performance compared to fully centralized settings (0.97 AUPRC) -- please see Table~\ref{tab:results}, which provides a concise summary of \SYSNAME's performance for different numbers of participants and privacy budgets.
However, this drop is due to the use of differentially private algorithms rather than the federated nature of \SYSNAME, as when we experiment with lower privacy settings, we quickly approach the fully centralized model's performance (e.g., 0.94 AUPRC with $\varepsilon = 10$).

This suggests there should be ample room to improve \SYSNAME's performance.
First, the effectiveness of inference attacks drops with large numbers of participants~\cite{melis2019exploiting}; thus, \SYSNAME does not necessarily need small values of $\varepsilon$.
Second, there are possible optimizations from the point of view of DP through the use of newer differentially private optimization algorithms like DP-Follow-The-Regularized-Leader~\cite{kairouz2021practical}, which performs better than DP-SGD at moderate to high levels of privacy. 
Finally, more advanced neural network architectures and performing feature selection on top of feature pre-processing, etc., can also be experimented with to achieve higher utility at the same level of privacy.
Since our main objective is to experiment with and assess the feasibility of a first-of-its-kind federated architecture for fingerprinting detection, we believe these improvements can be addressed in future work.

Moreover, we find that the overwhelming majority of misclassifications reducing AUPRC are false negatives.
These could be considered less problematic than false positives in the context of fingerprinting and, more importantly, could be reduced with access to larger datasets covering more fingerprinting scripts.
Also, as discussed, centralized crawls likely miss a number of fingerprinting scripts due to bot detection and user login walls, so their actual {\em recall} is likely much lower.

\descr{FL Bootstrap.} While \SYSNAME provides formal privacy guarantees, any real-world deployments of DP-FL face a number of practical challenges.
For instance, devices often join and drop out of FL systems in unpredictable ways, which makes it difficult to sample users in a truly random manner.
However, recent work~\cite{mcmahan2022federated} introduces new algorithms and privacy analysis techniques that mitigate this concern while providing comparable utility, at least in theory.
Since we use second-order methods (LBFGS) instead of the first-order methods (SGD) as in~\cite{mcmahan2022federated}, adapting their algorithms in a straightforward manner might potentially degrade \SYSNAME's model performance, and thus we leave this to future work.

\descr{Heuristics.} Finally, our ground truth relies on heuristics developed by FP-Inspector in 2019~\cite{iqbal2021fingerprinting}.
We did so since this is a well-established, high-precision heuristic.
However, the introduction of new APIs, such as WebGPU, and the imminent removal of third-party cookies from Chrome can potentially result in new types of fingerprinting being introduced.
Detecting these new scripts would require new heuristics to be defined.
Therefore, future work should continue to explore new avenues for fingerprinting and to define new high-precision heuristics.
\subsection{Acknowledgments}
This work has been supported by the National Science Scholarship (PhD) from the Agency for Science Technology and Research, Singapore, and a Google Research Faculty Award.
The authors also wish to thank Umar Iqbal for sharing their automated web crawl data and source code for FP-Inspector~\cite{iqbal2021fingerprinting}.

{\small
\bibliographystyle{abbrv}
%\bibliography{ref}

\begin{thebibliography}{10}

\bibitem{abadi2016deep}
M.~Abadi, A.~Chu, I.~Goodfellow, H.~B. McMahan, I.~Mironov, K.~Talwar, and
  L.~Zhang.
\newblock {Deep learning with differential privacy}.
\newblock In {\em ACM CCS}, 2016.

\bibitem{acar2014web}
G.~Acar, C.~Eubank, S.~Englehardt, M.~Juarez, A.~Narayanan, and C.~Diaz.
\newblock {The Web Never Forgets: Persistent Tracking Mechanisms in the Wild}.
\newblock In {\em {ACM CCS}}, 2014.

\bibitem{acar2013fpdetective}
G.~Acar, M.~Juarez, N.~Nikiforakis, C.~Diaz, S.~G{\"u}rses, F.~Piessens, and
  B.~Preneel.
\newblock {FPDetective: Dusting the Web for Fingerprinters}.
\newblock In {\em ACM CCS}, 2013.

\bibitem{adamic2002zipf}
L.~A. Adamic and B.~A. Huberman.
\newblock {Zipf's law and the Internet.}
\newblock {\em Glottometrics}, 3(1), 2002.

\bibitem{adnan2022federated}
M.~Adnan, S.~Kalra, J.~C. Cresswell, G.~W. Taylor, and H.~R. Tizhoosh.
\newblock {Federated learning and differential privacy for medical image
  analysis}.
\newblock {\em Scientific reports}, 12(1), 2022.

\bibitem{agarwal2021skellam}
N.~Agarwal, P.~Kairouz, and Z.~Liu.
\newblock {The Skellam Mechanism for Differentially Private Federated
  Learning}.
\newblock {\em NeurIPS}, 2021.

\bibitem{akhavani2021browserprint}
S.~A. Akhavani, J.~Jueckstock, J.~Su, A.~Kapravelos, E.~Kirda, and L.~Lu.
\newblock {Browserprint: An Analysis of the Impact of Browser Features on
  Fingerprintability and Web Privacy}.
\newblock In {\em Information Security Conference}, 2021.

\bibitem{alaca2016device}
F.~Alaca and P.~C. Van~Oorschot.
\newblock {Device fingerprinting for augmenting web authentication:
  classification and analysis of methods}.
\newblock In {\em {ACM CCS}}, 2016.

\bibitem{alexa2022}
Amazon.
\newblock {We will be retiring Alexa.com on May 1, 2022}.
\newblock
  \url{https://web.archive.org/web/20220102200605/https://support.alexa.com/hc/en-us/articles/4410503838999},
  2022.

\bibitem{amjad2023blocking}
A.~H. Amjad, Z.~Shafiq, and M.~A. Gulzar.
\newblock {Blocking JavaScript without Breaking the Web: An Empirical
  Investigation}.
\newblock {\em arXiv:2302.01182}, 2023.

\bibitem{brave2022}
P.~Arntz.
\newblock {Brave browser goes the extra mile to block third party cookies}.
\newblock
  \url{https://www.malwarebytes.com/blog/news/2022/03/brave-browser-goes-the-extra-mile-to-block-third-party-cookies},
  2022.

\bibitem{bagdasaryan2020backdoor}
E.~Bagdasaryan, A.~Veit, Y.~Hua, D.~Estrin, and V.~Shmatikov.
\newblock How to backdoor federated learning.
\newblock In {\em International Conference on Artificial Intelligence and
  Statistics}, 2020.

\bibitem{bahrami2021}
P.~N. Bahrami, U.~Iqbal, and Z.~Shafiq.
\newblock {FP-Radar: Longitudinal Measurement and Early Detection of Browser
  Fingerprinting}.
\newblock {\em PETS}, 2022, 2022.

\bibitem{batuwita2013class}
R.~Batuwita and V.~Palade.
\newblock {Class Imbalance Learning Methods for Support Vector Machines}.
\newblock {\em Imbalanced Learning: Foundations, Algorithms, and Applications},
  2013.

\bibitem{puppeteer}
Berstend.
\newblock {puppeteer-extra-plugin-stealth}.
\newblock \url{https://github.com/berstend/puppeteer-extra}, 2023.

\bibitem{bonawitz2019towards}
K.~Bonawitz, H.~Eichner, W.~Grieskamp, D.~Huba, A.~Ingerman, V.~Ivanov,
  C.~Kiddon, J.~Kone{\v{c}}n{\`y}, S.~Mazzocchi, B.~McMahan, T.~V. Overveldt,
  D.~Petrou, D.~Ramage, and R.~Jason.
\newblock {Towards Federated Learning at Scale: System Design}.
\newblock {\em Proceedings of Machine Learning and Systems}, 1, 2019.

\bibitem{bravebrowserfp}
Brave.
\newblock {Fingerprint Randomization}.
\newblock \url{https://brave.com/privacy-updates/3-fingerprint-randomization/},
  2020.

\bibitem{bursztein2016picasso}
E.~Bursztein, A.~Malyshev, T.~Pietraszek, and K.~Thomas.
\newblock {Picasso: Lightweight device class fingerprinting for web clients}.
\newblock In {\em {Workshop on Security and Privacy in Smartphones and Mobile
  Devices}}, 2016.

\bibitem{safaribrowserfp2018}
D.~Cameron.
\newblock {Apple Declares War on 'Browser Fingerprinting,' the Sneaky Tactic
  That Tracks You in Incognito Mode}.
\newblock
  \url{https://gizmodo.com/apple-declares-war-on-browser-fingerprinting-the-sneak-1826549108},
  2018.

\bibitem{crux}
Chrome.
\newblock {Chrome User Experience Report}.
\newblock \url{https://developer.chrome.com/docs/crux/}.

\bibitem{clauset2009power}
A.~Clauset, C.~R. Shalizi, and M.~E. Newman.
\newblock {Power-Law Distributions in Empirical Data}.
\newblock {\em SIAM Review}, 51(4), 2009.

\bibitem{das2018web}
A.~Das, G.~Acar, N.~Borisov, and A.~Pradeep.
\newblock {The Web's Sixth Sense: A Study of Scripts Accessing Smartphone
  Sensors}.
\newblock In {\em {ACM CCS}}, 2018.

\bibitem{nhsfb}
S.~Das.
\newblock {NHS data breach: trusts shared patient details with Facebook without
  consent}.
\newblock
  \url{https://www.theguardian.com/society/2023/may/27/nhs-data-breach-trusts-shared-patient-details-with-facebook-meta-without-consent?CMP=Share_iOSApp_Other},
  2022.

\bibitem{nhs2023}
S.~Das.
\newblock {NHS data breach: trusts shared patient details with Facebook without
  consent}.
\newblock
  \url{https://www.theguardian.com/society/2023/may/27/nhs-data-breach-trusts-shared-patient-details-with-facebook-meta-without-consent},
  2023.

\bibitem{disconnect2018}
Disconnect.
\newblock {Disconnect defends the digital you}.
\newblock \url{https://disconnect.me}, 2018.

\bibitem{dwork2014algorithmic}
C.~Dwork and A.~Roth.
\newblock {The Algorithmic Foundations of Differential Privacy}.
\newblock {\em Foundations and Trends in Theoretical Computer Science}, 2014.

\bibitem{easylist}
EasyList.
\newblock \url{https://easylist-downloads.adblockplus.org/}.

\bibitem{easyprivacy}
EasyList.
\newblock {EasyPrivacy}.
\newblock \url{https://easylist.to/easylist/easyprivacy.txt}.

\bibitem{eckersley2010unique}
P.~Eckersley.
\newblock {How Unique Is Your Web Browser?}
\newblock {\em PETS}, 2010.

\bibitem{privacybadger}
{EFF}.
\newblock {{PrivacyBadger}}.
\newblock
  \url{https://github.com/EFForg/privacybadgerfirefox/blob/master/data/cookieblocklist.txt},
  2023.

\bibitem{firefoxbrowserfp2020}
S.~Englehardt.
\newblock {Firefox 72 blocks third-party fingerprinting resources}.
\newblock
  \url{https://blog.mozilla.org/security/2020/01/07/firefox-72-fingerprinting/},
  2020.

\bibitem{englehardt2016online}
S.~Englehardt and A.~Narayanan.
\newblock {Online tracking: A 1-million-site Measurement and Analysis}.
\newblock In {\em {ACM CCS}}, 2016.

\bibitem{fingerprintpro}
Fingerprint.
\newblock {Bot Detection Guide}.
\newblock
  \url{https://dev.fingerprint.com/docs/bot-detection-quick-start-guide}.

\bibitem{galvez2020less}
R.~G{\'a}lvez, V.~Moonsamy, and C.~Diaz.
\newblock {Less is More: A privacy-respecting Android malware classifier using
  federated learning}.
\newblock {\em arXiv:2007.08319}, 2020.

\bibitem{geyer2017differentially}
R.~C. Geyer, T.~Klein, and M.~Nabi.
\newblock {Differentially private federated learning: A client level
  perspective}.
\newblock {\em arXiv:1712.07557}, 2017.

\bibitem{hartmann2023distributed}
F.~Hartmann and P.~Kairouz.
\newblock {Distributed differential privacy for federated learning}.
\newblock
  \url{https://ai.googleblog.com/2023/03/distributed-differential-privacy-for.html},
  2023.

\bibitem{ikram2017towards}
M.~Ikram, H.~J. Asghar, M.~A. Kaafar, B.~Krishnamurthy, and A.~Mahanti.
\newblock {Towards Seamless Tracking-Free Web: Improved Detection of Trackers
  via One-class Learning}.
\newblock {\em PETS}, 2017.

\bibitem{iovationfraud2019}
Iovation.
\newblock {Iovation Fraud Protection}.
\newblock
  \url{https://web.archive.org/web/20191130164107/\\https://www.iovation.com/fraudforce-fraud-detection-prevention},
  2019.

\bibitem{iqbal2021fingerprinting}
U.~Iqbal, S.~Englehardt, and Z.~Shafiq.
\newblock {Fingerprinting the fingerprinters: Learning to detect browser
  fingerprinting behaviors}.
\newblock In {\em {IEEE S\&P}}, 2021.

\bibitem{jshelter}
JShelter.
\newblock \url{https://jshelter.org/}.

\bibitem{jordan2021towards}
J.~Jueckstock, S.~Sarker, P.~Snyder, A.~Beggs, P.~Papadopoulos, M.~Varvello,
  B.~Livshits, and A.~Kapravelos.
\newblock {Towards Realistic and Reproducible Web Crawl Measurements}.
\newblock In {\em WWW}, 2021.

\bibitem{kairouz2021distributed}
P.~Kairouz, Z.~Liu, and T.~Steinke.
\newblock {The Distributed Discrete Gaussian Mechanism for Federated Learning
  with Secure Aggregation}.
\newblock In {\em ICML}, 2021.

\bibitem{kairouz2021practical}
P.~Kairouz, B.~Mcmahan, S.~Song, O.~Thakkar, A.~Thakurta, and Z.~Xu.
\newblock {Practical and Private (Deep) Learning Without Sampling or
  Shuffling}.
\newblock In {\em ICML}, 2021.

\bibitem{kairouz2021advances}
P.~Kairouz, H.~B. McMahan, B.~Avent, A.~Bellet, M.~Bennis, A.~N. Bhagoji,
  K.~Bonawitz, Z.~Charles, G.~Cormode, R.~Cummings, et~al.
\newblock {Advances and Open Problems in Federated Learning}.
\newblock {\em Foundations and Trends in Machine Learning}, 14(1--2), 2021.

\bibitem{kairouz2015composition}
P.~Kairouz, S.~Oh, and P.~Viswanath.
\newblock The composition theorem for differential privacy.
\newblock In {\em ICML}, 2015.

\bibitem{kang2019incentive}
J.~Kang, Z.~Xiong, D.~Niyato, S.~Xie, and J.~Zhang.
\newblock {Incentive Mechanism for Reliable Federated Learning: A Joint
  Optimization Approach to Combining Reputation and Contract Theory}.
\newblock {\em IEEE Internet of Things Journal}, 6(6), 2019.

\bibitem{krishnamurthy2009privacy}
B.~Krishnamurthy and C.~Wills.
\newblock {Privacy diffusion on the web: a longitudinal perspective}.
\newblock In {\em WWW}, 2009.

\bibitem{torbrowserfp}
P.~Laperdrix.
\newblock {Browser Fingerprinting: An Introduction and the Challenges Ahead}.
\newblock
  \url{https://blog.torproject.org/browser-fingerprinting-introduction-and-challenges-ahead/},
  2019.

\bibitem{laperdrix2019morellian}
P.~Laperdrix, G.~Avoine, B.~Baudry, and N.~Nikiforakis.
\newblock {Morellian analysis for browsers: Making web authentication stronger
  with canvas fingerprinting}.
\newblock In {\em {Detection of Intrusions and Malware, and Vulnerability
  Assessment}}, 2019.

\bibitem{chrome2023}
R.~Lawler.
\newblock {Google’s turning off third-party cookies for 1 percent of Chrome
  users early next year}.
\newblock
  \url{https://www.theverge.com/2023/5/18/23728263/google-chrome-ad-tracking-third-party-cookies-privacy-sandbox},
  2023.

\bibitem{li2020deepfed}
B.~Li, Y.~Wu, J.~Song, R.~Lu, T.~Li, and L.~Zhao.
\newblock {DeepFed: Federated deep learning for intrusion detection in
  industrial cyber--physical systems}.
\newblock {\em IEEE Transactions on Industrial Informatics}, 17(8), 2020.

\bibitem{liu2020deep}
Y.~Liu, S.~Garg, J.~Nie, Y.~Zhang, Z.~Xiong, J.~Kang, and M.~S. Hossain.
\newblock {Deep anomaly detection for time-series data in industrial IoT: A
  communication-efficient on-device federated learning approach}.
\newblock {\em IEEE Internet of Things Journal}, 8(8), 2020.

\bibitem{relix2018}
I.~Lunden.
\newblock {Relx acquires ThreatMetrix for \$817M to ramp up in risk-based
  authentication}.
\newblock
  \url{https://techcrunch.com/2018/01/29/relx-threatmetrix-risk-authentication-lexisnexis/?guccounter=1},
  2018.

\bibitem{maddock2022federated}
S.~Maddock, G.~Cormode, T.~Wang, C.~Maple, and S.~Jha.
\newblock {Federated Boosted Decision Trees with Differential Privacy}.
\newblock In {\em ACM CCS}, 2022.

\bibitem{mcmahan2022federated}
B.~McMahan and A.~Thakurta.
\newblock {Federated learning with formal differential privacy guarantees}.
\newblock {\em Google AI Blog}, 2022.

\bibitem{mcmahan2016federated}
H.~B. McMahan, E.~Moore, D.~Ramage, and B.~A. y~Arcas.
\newblock {Federated learning of deep networks using model averaging}.
\newblock {\em arXiv:1602.05629}, 2016.

\bibitem{mcmahan2018learning}
H.~B. McMahan, D.~Ramage, K.~Talwar, and L.~Zhang.
\newblock {Learning Differentially Private Recurrent Language Models}.
\newblock In {\em ICLR}, 2018.

\bibitem{melis2019exploiting}
L.~Melis, C.~Song, E.~De~Cristofaro, and V.~Shmatikov.
\newblock {Exploiting Unintended Feature Leakage in Collaborative Learning}.
\newblock In {\em IEEE S\&P}, 2019.

\bibitem{mothukuri2021federated}
V.~Mothukuri, P.~Khare, R.~M. Parizi, S.~Pouriyeh, A.~Dehghantanha, and
  G.~Srivastava.
\newblock {Federated-Learning-Based Anomaly Detection for IoT Security
  Attacks}.
\newblock {\em IEEE Internet of Things Journal}, 9(4), 2021.

\bibitem{webgpu}
Mozilla.
\newblock {WebGPU API}.
\newblock \url{https://developer.mozilla.org/en-US/docs/Web/API/WebGPU_API},
  2023.

\bibitem{webrtc}
Mozilla.
\newblock {WebRTC API}.
\newblock \url{https://developer.mozilla.org/en-US/docs/Web/API/WebRTC_API},
  2023.

\bibitem{canvasdefender}
MultiLogin.
\newblock {Canvas Defender}.
\newblock \url{https://multilogin.com/canvas-defender/}.

\bibitem{naseri2022cerberus}
M.~Naseri, Y.~Han, E.~Mariconti, Y.~Shen, G.~Stringhini, and E.~De~Cristofaro.
\newblock {Cerberus: Exploring Federated Prediction of Security Events}.
\newblock In {\em ACM CCS}, 2022.

\bibitem{naseri2022local}
M.~Naseri, J.~Hayes, and E.~De~Cristofaro.
\newblock {Local and Central Differential Privacy for Robustness and Privacy in
  Federated Learning}.
\newblock In {\em NDSS}, 2022.

\bibitem{nasr2019comprehensive}
M.~Nasr, R.~Shokri, and A.~Houmansadr.
\newblock {Comprehensive Privacy Analysis of Deep Learning: Passive and Active
  White-box Inference Attacks against Centralized and Federated Learning}.
\newblock In {\em IEEE S\&P}, 2019.

\bibitem{ngan2022nowhere}
R.~Ngan, S.~Konkimalla, and Z.~Shafiq.
\newblock {Nowhere to Hide: Detecting Obfuscated Fingerprinting Scripts}.
\newblock {\em arXiv:2206.13599}, 2022.

\bibitem{pihur2018differentially}
V.~Pihur, A.~Korolova, F.~Liu, S.~Sankuratripati, M.~Yung, D.~Huang, and
  R.~Zeng.
\newblock {Differentially-private ``draw and discard'' machine learning}.
\newblock In {\em CSCML}, 2022.

\bibitem{vasyl2022}
V.~Pihur, A.~Korolova, F.~Liu, S.~Sankuratripati, M.~Yung, D.~Huang, and
  R.~Zeng.
\newblock {Differentially-Private ``Draw and Discard'' Machine Learning:
  Training Distributed Model from Enormous Crowds}.
\newblock In {\em Cyber Security, Cryptology, and Machine Learning Symposium},
  2022.

\bibitem{victor2019tranco}
V.~L. Pochat, T.~van Goethem, S.~Tajalizadehkhoob, M.~Korczynski, and
  W.~Joosen.
\newblock {Tranco: A Research-Oriented Top Sites Ranking Hardened Against
  Manipulation}.
\newblock In {\em NDSS}, 2019.

\bibitem{pugliese2020long}
G.~Pugliese, C.~Riess, F.~Gassmann, and Z.~Benenson.
\newblock {Long-Term Observation on Browser Fingerprinting: Users' Trackability
  and Perspective}.
\newblock {\em PETS}, 2020.

\bibitem{rack2023jack}
J.~Rack and C.-A. Staicu.
\newblock {Jack-in-the-box: An Empirical Study of JavaScript Bundling on the
  Web and its Security Implications}.
\newblock {\em CCS}, 2023.

\bibitem{rey2022federated}
V.~Rey, P.~M.~S. S{\'a}nchez, A.~H. Celdr{\'a}n, and G.~Bovet.
\newblock {Federated learning for malware detection in iot devices}.
\newblock {\em Computer Networks}, 204, 2022.

\bibitem{roesner2012detecting}
F.~Roesner, T.~Kohno, and D.~Wetherall.
\newblock {Detecting and Defending Against Third-Party Tracking on the Web}.
\newblock In {\em USENIX}, 2012.

\bibitem{ruth2022toppling}
K.~Ruth, D.~Kumar, B.~Wang, L.~Valenta, and Z.~Durumeric.
\newblock Toppling top lists: Evaluating the accuracy of popular website lists.
\newblock In {\em ACM IMC}, 2022.

\bibitem{fingerprinting2019}
J.~Schuh.
\newblock {Building a more private web}.
\newblock
  \url{https://www.blog.google/products/chrome/building-a-more-private-web/},
  2019.

\bibitem{sun2021ldp}
L.~Sun, J.~Qian, and X.~Chen.
\newblock {LDP-FL: Practical Private Aggregation in Federated Learning with
  Local Differential Privacy}.
\newblock In {\em IJCAI}, 2021.

\bibitem{truex2020ldp}
S.~Truex, L.~Liu, K.-H. Chow, M.~E. Gursoy, and W.~Wei.
\newblock {LDP-Fed: Federated learning with local differential privacy}.
\newblock In {\em ACM International Workshop on Edge Systems, Analytics and
  Networking}, 2020.

\bibitem{w3cbrowserfp}
W3C.
\newblock {Mitigating Browser Fingerprinting in Web Specifications}.
\newblock \url{https://w3c.github.io/fingerprinting-guidance/}, 2021.

\bibitem{thirdpartycookies}
W3C.
\newblock {Improving the web without third-party cookies}.
\newblock \url{https://www.w3.org/2001/tag/doc/web-without-3p-cookies/}, 2023.

\bibitem{webgpuw3c}
W3C.
\newblock {WebGPU}.
\newblock \url{https://www.w3.org/TR/webgpu/}, 2023.

\bibitem{privacyBudget2022}
A.~White.
\newblock {Google’s turning off third-party cookies for 1 percent of Chrome
  users early next year}.
\newblock
  \url{https://developer.chrome.com/docs/privacy-sandbox/privacy-budget/},
  2022.

\bibitem{bounce2020}
J.~Wilander.
\newblock {Bounce Tracking Protection}.
\newblock \url{https://github.com/privacycg/proposals/issues/6}, 2020.

\bibitem{safari2020}
J.~Wilander.
\newblock {Full Third-Party Cookie Blocking and More}.
\newblock
  \url{https://webkit.org/blog/10218/full-third-party-cookie-blocking-and-more/},
  2020.

\bibitem{firefox2019}
M.~Wood.
\newblock {Todays Firefox Blocks Third-Party Tracking Cookies and Cryptomining
  by Default}.
\newblock
  \url{https://blog.mozilla.org/en/products/firefox/todays-firefox-blocks-third-party-tracking-cookies-and-cryptomining-by-default/},
  2019.

\bibitem{cloudflarebot}
ZenRows.
\newblock {How to Bypass Cloudflare in 2023: The 8 Best Methods}.
\newblock \url{https://www.zenrows.com/blog/bypass-cloudflare}, 2023.

\end{thebibliography}

}

\appendices
  
\section{Missed Scripts}
\label{app:missed_scripts}
As mentioned, since centralized crawlers cannot replicate real human interactions, they might miss scripts on webpages protected by bot detectors, CAPTCHAs, and webpages that require user login.
One specific example is available from \url{https://docs.google.com/document/d/1luaU5I8oU8wVt-QFy5muGjM47GkW57I8rO8n8dcyX-8/}.\footnote{In the spirit of responsible disclosure, we omit the domain name where the script has been found along with any identification detail.}
This script, found in the wild during our top 300 crawl (see Section~\ref{subsec:preproc}), enumerates various device properties, i.e., screen size, language, device timezone, plugins installed, and WebGL configuration, on top of performing canvas and audio fingerprinting.

Even though we use a bot detection evasion tool\footnote{https://www.npmjs.com/package/puppeteer-extra-plugin-stealth}, our automated crawler was detected as a bot and prevented from visiting the website, thus missing the script.
On the other hand, our manual crawl with real user interaction was able to visit the website and flag the script as fingerprinting.

\section{Non-IIDness score}
\label{app:noniidness}
To quantify the Non-IIDness of the distributions tested, we follow an approach inspired by previous work~\cite{naseri2022cerberus}, which introduces and uses the so-called {\em Non-IIDness score.}
This calculates the average pairwise distance, expressed in terms of Kullback–Leibler (KL) divergence, between the distribution of classes among participants. For instance, when the training data is IID distributed, the distribution of classes among participants is roughly similar, which leads to a small average distance and a small Non-IIDness score.
However, when the training data is unevenly distributed, the distribution of classes is starkly different, resulting in a high average pairwise distance between distributions and thus high Non-IIDness score.

To use it in our setting, we need to introduce some modifications. 
First, while~\cite{naseri2022cerberus} calculates the average distance across all participants, the number of participants in our setting is much greater than in theirs (we have millions of users, whereas they only had a few thousand).
Thus, we cannot enumerate (possibly) trillions of pairwise combinations.
Rather, we calculate the average over a sample of 1000 participants (corresponding to roughly 1M pairwise combinations).

Next, since browser fingerprinting is a heavily class-imbalanced problem, the distribution of fingerprinting scripts in participants does not change the overall distribution of scripts significantly.
Therefore, in this work, we calculate the distance between the distribution of only fingerprinting scripts by type (excluding non-fingerprinting scripts).

Finally, since scripts can contain multiple types of fingerprinting behavior, the distribution of fingerprinting scripts consists of not just each individual type of fingerprinting but all combinations of types of fingerprinting as well (e.g., Canvas, Canvas Font, Audio, WebRTC, Canvas and Canvas Font, Canvas and Audio, etc.).

\end{document}